\RequirePackage{fix-cm}
\documentclass[smallextended]{svjour3}
\usepackage[14pt]{extsizes}
\smartqed   
\usepackage{graphicx}
\usepackage{amsmath}
\usepackage{amsfonts}
\usepackage{amssymb}
\usepackage{makeidx}
\usepackage{float}
\usepackage{subfig}
\usepackage{enumerate} 
\usepackage{fancyhdr}
\usepackage{pdfpages}
\usepackage{cite}
\usepackage{changes}
\usepackage{color}
\usepackage{soul}
\usepackage{verbatim}

\makeatletter
\renewcommand\@biblabel[1]{}
\makeatother

\begin{document}

\title{Tsunami hydrodynamic force on a building using a SPH real scale numerical simulation}


\author{Jaime Klapp$^{1,*}$     \and
        Omar S. Areu-Rangel$^{2}$    \and
        Marcela Cruchaga$^{3}$     \and
        Rafael Ar\'anguiz$^{4,5}$    \and
        Rosanna Bonasia$^{6}$    \and
        Mauricio Godoy Seura$^{7}$ 
}


\institute{$^{1,*}$Instituto Nacional de Investigaciones Nucleares - Departamento de F\'isica, M\'exico
               \\
           \email{jaime.klapp@inin.gob.mx}
           \\
          $^{2}$Tecnol\'ogico Nacional de M\'exico/Campus Pachuca  - Departamento de Ingenier\'ia Civil, M\'exico
          \\ 
          $^{3}$Universidad de Santiago de Chile - Depto. Ingenier\'ia Mec\'anica, Chile 
          \\
          $^{4}$ Universidad Cat\'olica de la Sant\'isima Concepci\'on - Departamento de Ingenier\'ia Civil, Chile  
          \\
          $^{5}$ National Research Center for Integrated Natural Disaster and Management, CIGIDEN, Chile  
          \\
          $^{6}$ CONACYT - Secci\'on de Estudios de Posgrado e Investigaci\'on, Instituto Politécnico Nacional, ESIA Zacatenco, Mexico City, Mexico  
         \\
             $^{7}$ Universidad de La Serena - Depto. de Ingenier\'ia Mec\'anica, Chile 
}

\date{Received: date / Accepted: date}

\maketitle

\begin{abstract}
One of the most important aspects in tsunami studies is the wave behavior when it approaches the coast. Information on physical parameters that characterize waves is often limited because of the difficulties in achieving accurate measurements at the time of the event. The impact of a tsunami on the coast is governed by nonlinear physics such as turbulence with spatial and temporal variability. The use of the Smoothed Particle Hydrodynamic method (SPH) presents advantages over models based on two-dimensional Shallow Waters Equations (SWE), because the assumed vertical velocity simplifies hydrodynamics in two dimensions. 
The study presented here reports numerical SPH simulations of the tsunami event occurred in Coquimbo (Chile) on September 16 of 2015. On the basis of the reconstruction of the physical parameters that characterized this event (flow velocities, direction and water elevations), calibrated by a reference rodel, force values on buildings located on the study coast were numerically calculated, and compared with an estimate of the Chilean Structural Design Standard. 
Finally, discussion and conclusions of the comparison of both methodologies are presented, including an influence analysis of the topographical detail of the model in the estimation of hydrodynamic forces.
\end{abstract}

\section{Introduction}
\label{intro}
Tsunamis are natural phenomena which are generated by megathrust earthquakes in subduction zones, such as the 2004 Indian Ocean, 2010 Chile and 2011 Great East Japan tsunamis. These phenomena have devastated different vulnerable zones throughout history causing great human, economic and natural losses. In addition, it is difficult to predict the occurrence of a tsunami, and there are only few minutes to issue an evacuation warning once a tsunami has been detected, hence the definition of tsunami evacuation areas has an important role on saving lives, so that people can evacuate immediately after the tsunami warning has been issued. However, in extensive plain coasts, where high grounds are not present, the construction of tsunami shelters become important for vertical evacuation.  

Structural designs of buildings and other coastal structures that could be subjected to tsunami loads has been investigated in recent years. As a matter of fact, the Federal Emergency Management Agency (FEMA) published the Guidelines for Design of Structures for Vertical Evacuation from Tsunamis (FEMA 2008). In Addition, the 2016 edition of ASCE Minimum Design Loads and Associated Criteria for Buildings and Other Structures incorporated means for determining tsunami loads for general structural design (ASCE, 2016). In a similar manner, the Chilean National Institute for Normalization (INN) published the "Structural Design-Building in risk areas of flooding due to tsunami or seiche" (NCh3363, 2015). The previously mentioned guidelines describe the methodologies to estimate different tsunami loads such as hydrostatic, hydrodynamic, bouyancy, and impact forces due to debris, etc. For instance, the hydrodynamic force on a structure is computed from an analytical expression given in terms of the maximum momentum flux. This variable may be computed either from the flow velocity and inundation depth given by a numerical simulation or from an approximation based on the maximum runup (NCh3363, 2015). Even though the former uses high resolution grids for numerical simulations, the presence of structures is not incorporated. Therefore, changes in tsunami flow due to the presence of structures is not included in the analysis.   

More sophisticated models, such as the Smoothed Particle Hydrodynamics (SPH) model, have been used to simulate coastal flows. Some studies have used this model to perform flow simulations involving complex bathymetries (De Leffe et al., 2010) and large waves mitigation by different types of dikes (Crespo et al., 2007). In addition, recent works have investigated tsunami forces on structures. Wei \& Dalrymple (2016, 2015) and St-Germain et al. (2013) emphasized the study of hydrodynamic forces on structures, such as piles and bridges of different sections and orientation against flow. These works compared numerical SPH models with experimental models, showing a good agreement in the results of both models. However, they used some simplifications that mainly involved scaled models (no larger than 30m for numerical models) and flow conditions imposed by dam breaks. 

This paper estimates the hydrodynamic force on a building  based on the Sept 16, 2015 Coquimbo tsunami and compares the results with analytical expression given by the Chilean Standard NCh3363. The hydrodynamic force is computed by means of the SPH code DualSPHysics (Crespo et al., 2015). The inflow conditions (tsunami amplitude and flow velocity) are taken from a classical numerical simulation (Ar\'anguiz, R. et al., 2017) validated with tide gauges and field data.  The main contribution of this work is the real-scale tsunami simulation using a SPH model, as well as a sensitivity analysis of the influence of the topography resolution.

The paper is organized as follows: Section 2 presents the mathematical formulation of the SPH model; section 3 gives a description of the studied area; section 4 describes the methodology, which includes definition of the domain and, validation with respect to the reference model; section 5 presents the results; and, finally, main conclusions are discussed in section 6.
\section{DualSPHysics model}
\label{sec:1}
Numerical simulations of the 2015 tsunami in Coquimbo were performed using the DualSHPhysics code (Crespo et al., 2015), an open-source code that implements a SPH mesh-free Lagrangian model in Graphic Processing Units (GPUs). The code solves the Navier Stokes equations, i.e, the momentum conservation equation and the continuity equation,

\begin{equation}\label{eq:din1}
\frac{\  d v}{d t} = - \frac{\  1}{\rho} \triangledown p + g + \Gamma \, , 
\end{equation}	
\begin{equation}\label{eq:conti2}
\frac{\  d \rho}{d t} + \rho \triangledown \cdot v =0 \, ,
\end{equation}	
where $\rho$ refers to density, $t$ is the time, $v$ is the velocity, $p$ is the pressure, $g$ is the gravity acceleration and corresponds to a body force, and $\Gamma$ refers to dissipative terms. 

The SPH method calculates the value of any property of the fluid and its derivatives by interpolating the values of neighboring particles within its environment using an interpolation function called Kernel ($W$). The interpolation of the function $F(r)$ defined in $r'$  is $F(r)= \int {F(r' ) W(r' , h) d r' }$ (Gingold \& Monaghan, 1977),
where $r$ is the position of particle $a$ to be interpolated, $r'$ refers to the position of particle $b$ used for interpolation, $h$ is the smoothing length (distance range used for the interpolation) and $W(r-r', h)$ is the weighting function known as Kernel. The Lagrangian approximation form of such equation is
\begin{equation}\label{chen}	
F(r_a)= \sum_b {m_b \frac{\ F(r_b) }{\rho_b} W(r_a- r_b, h)} \, ,
\end{equation}	
where the subscript $b$ refers to a particle within the integration domain (with distance range $h$) of the Kernel function in the environment of the reference particle $a$, $m_b$ and $\rho_b$ are the mass and the density of particle $b$, $r_a$ is the position of particle $a$ and $W(r- r_b, h)$ is the weight function of particle $b$ with respect to particle $a$. The subscript $b$ ranges from $1$ to $N_p$, the latter being the total number of neighboring particles within the integration domain.

In the context of a SPH method based on a weakly compressible formulation, where for any fluid particle the mass remains constant and only the density fluctuates, the discrete form of Eqs. \ref{eq:din1} and \ref{eq:conti2} are  written as (Crespo et al., 2015) 
  \begin{equation}\label{continuidad1}	
  \frac{d \rho_a}{d t}=\sum_{b}^{} m_b \nu_{ab} \cdot \bigtriangledown_a W_{ab} \, .        
  \end{equation}	  	
  
  \begin{equation}\label{eq:din11}
  \frac{d v_a}{d t} = - \sum_b m_b \left(\frac{\  P_b + P_a}{\rho_b \cdot \rho_a} + \Pi_{ab} \right) \triangledown_a W_{ab}+g \, ,
  \end{equation}			
  where $P_k$ and $\rho_k$ are the pressure and density of the $k$ particle and the viscous term $\Pi_{ab}$ is defined by an artificial viscosity scheme developed by Monaghan (1992) as
  
  \begin{equation}\label{visco1}	
  \Pi_{ab} = \left \{ \begin{matrix} \frac{- \alpha C_{ab} \mu_{ab}}{ \rho_{ab}} & \nu_{ab} \cdot r_{ab} < 0
  \\ 0 & \nu_{ab} \cdot r_{ab} > 0 \end{matrix}\right.
  \end{equation}	             
  where $r_{ab}=r_a-r_b$, $\nu_{ab}=\nu_a-\nu_b$, with $r_i$ and $\nu_i$ equal to the position and velocity of the particle $i$. $\mu_{ab}= h \nu_{ab} \cdot r_{ab}/({r_{ab}}^2+ \eta^2)$, $C_{ab}=0.5(C_a+C_b)$ is the average of the speed of sound, $\eta^2=0.01 h^2$ and $\alpha$ is the dissipation coefficient.
  
  DualSPHysics reproduces structures by means of a fixed-position particle array. The dummy acceleration in the fixed particles can be calculated by solving the interpolation of the neighboring fluid particles with the Eq. \ref{eq:din11}. Then, the total force on fixed particles can be calculated solving the sum of the acceleration multiplied by the mass of each particle (Crespo et al., 2015) as follows
  \begin{equation}\label{eq:dinF}
  F = m  \sum_a \frac{\  d v_a}{d t} \, .
  \end{equation}	
  
\section{Study area}
\label{sec:3}
 Chile is extremely exposed to earthquakes because of its location in the Pacific Ring of Fire and the subduction of the Nazca plate under the South American plate. This makes it susceptible to tsunamis originated by the abrupt movements of the tectonic plates. Forces due to tsunami waves are capable of moving large objects such as ships, containers and even buildings over a highly populated area.

\begin{figure}
 	\centering
 	\includegraphics[width=0.8\textwidth]{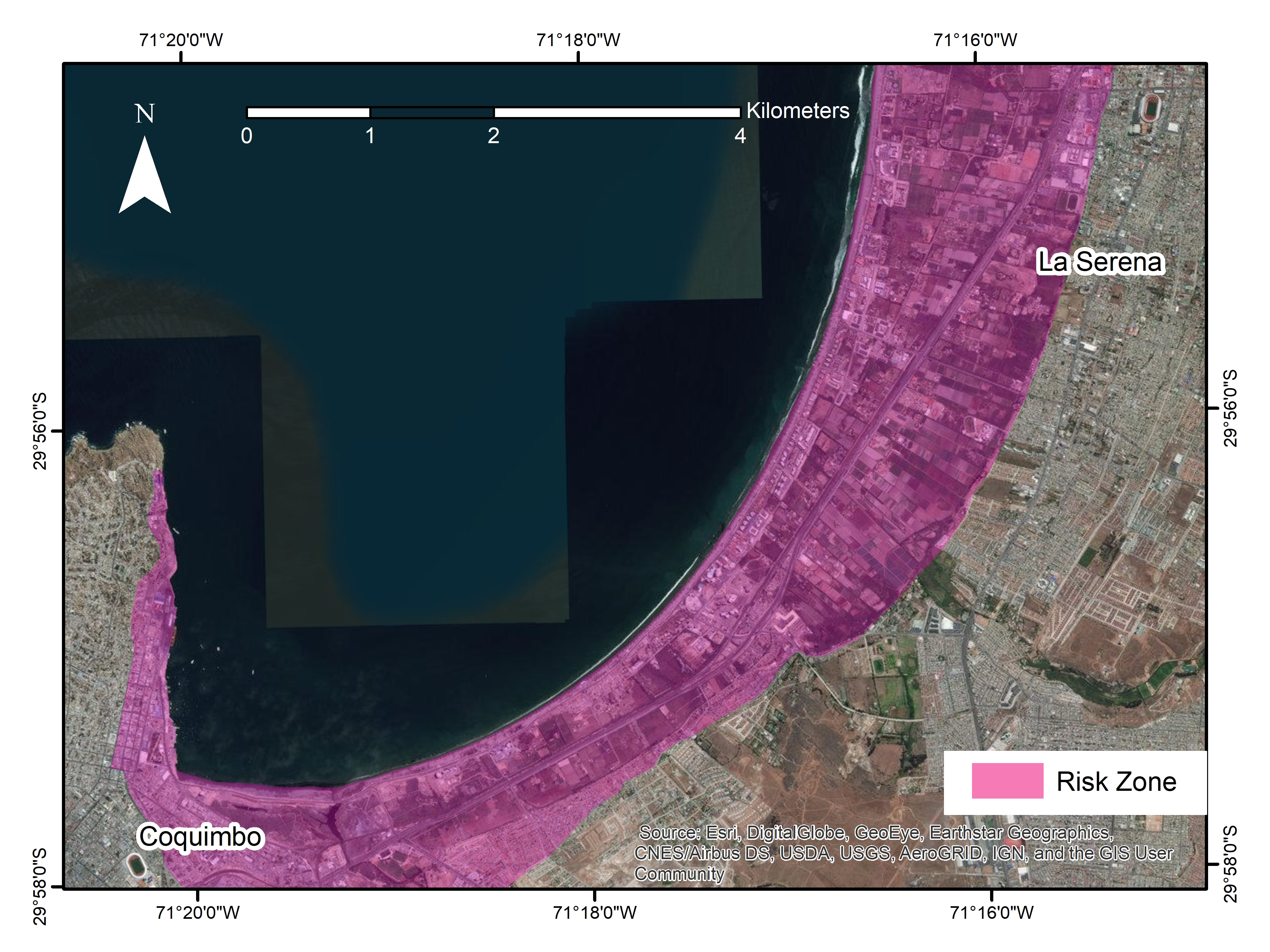}
 	\caption{Tsunami inundation hazard map of La Serena and Coquimbo, under the safe line.}
 	\label{fig:serena11}	 
 \end{figure}

 Coquimbo is located on the coast of northern Chile (29.96W, 71.34S). The city has been affected by large tsunamis in the past. The 1922 tsunami has been one of the most devastating events (Solovien and Go, 1984; Carvajal et al. 2016). The tsunami hit Coquimbo two hours after the earthquake, and the maximum inundation height at the southern shore of the bay reached up to 7 m above mean sea level (Soloviev and Go, 1984). Figure \ref{fig:serena11} shows the tsunami inundation hazard map of Coquimbo, which was defined with the use of numerical simulation results of the probable extreme event (SHOA, 2015). All people living within the coloured area should evacuate in case of a tsunami warning.

 The most recent tsunami that affected Coquimbo occurred in 2015. It caused significant inundations that produced serious damages to buildings, infrastructures and communication, as well as the loss of ten lives (Aranguiz et al., 2016). The tsunami was generated by an earthquake of magnitude Mw 8.3, and reached the coastal zones a few minutes after the earthquake (Aranguiz et al., 2016). The Pacific Tsunami Warning Center (PTWC) and the Hydrographic and Oceanographic Service of the Chilean Navy (SHOA) were able to issue a tsunami warning message a few minutes after the earthquake. However, this did not avoid the damage that the event caused (ONEMI, 2015).
 
\begin{figure*}
  	\centering
  	\subfloat[]{
  		\includegraphics[width=0.5\textwidth]{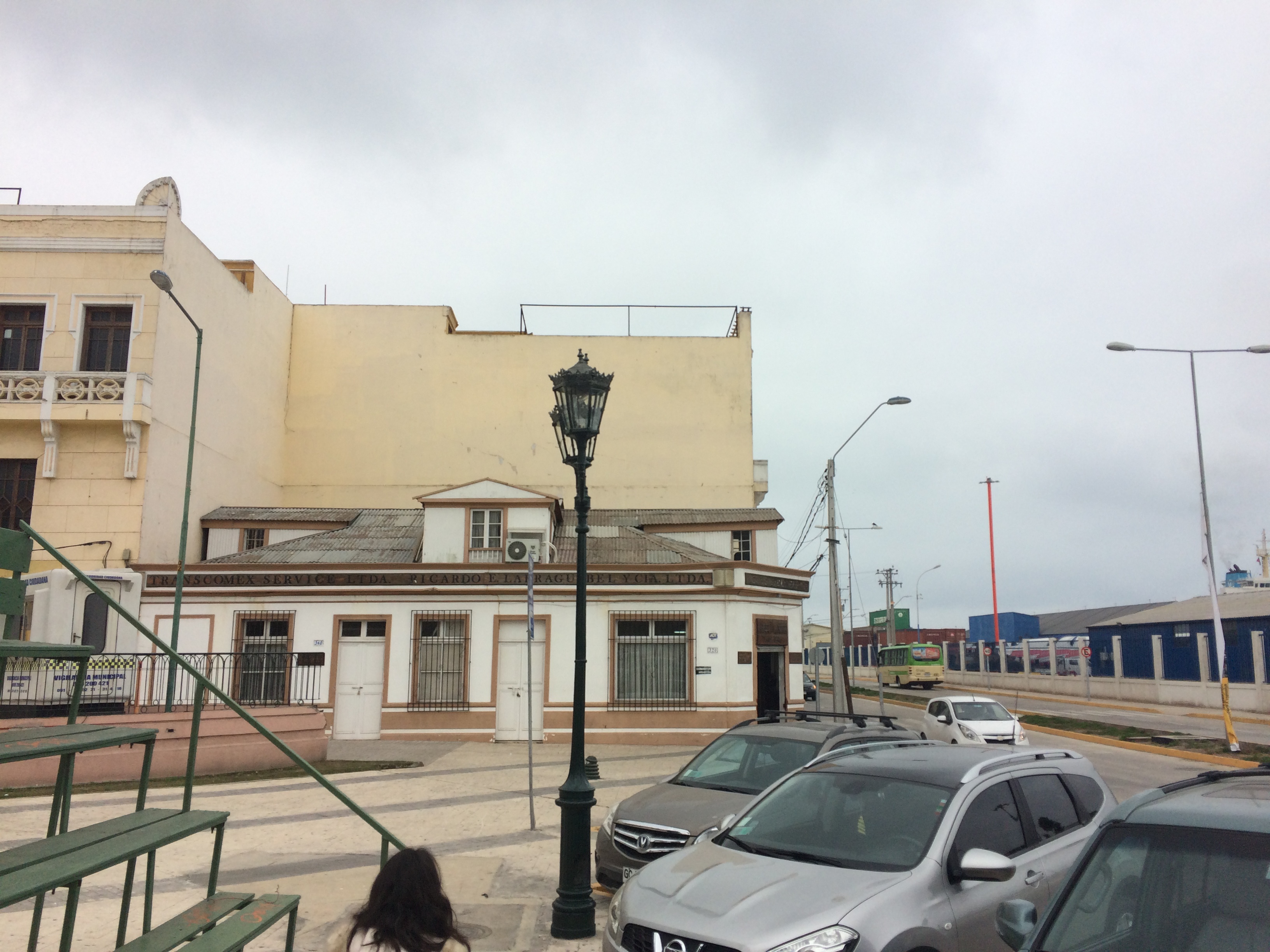}}
  	\subfloat[]{
  		\includegraphics[width=0.5\textwidth]{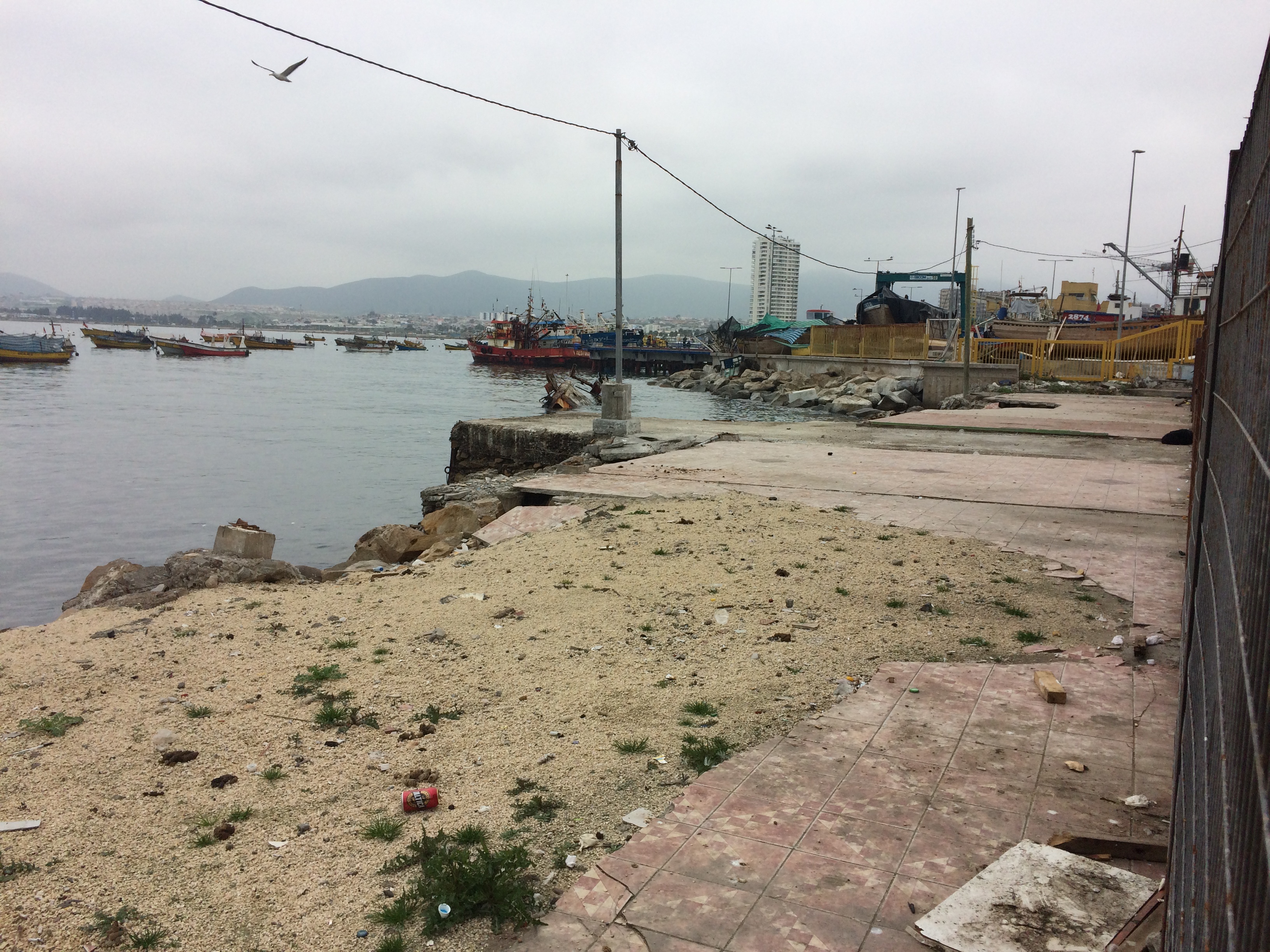}}\\
  	\subfloat[]{
  		\includegraphics[width=0.5\textwidth]{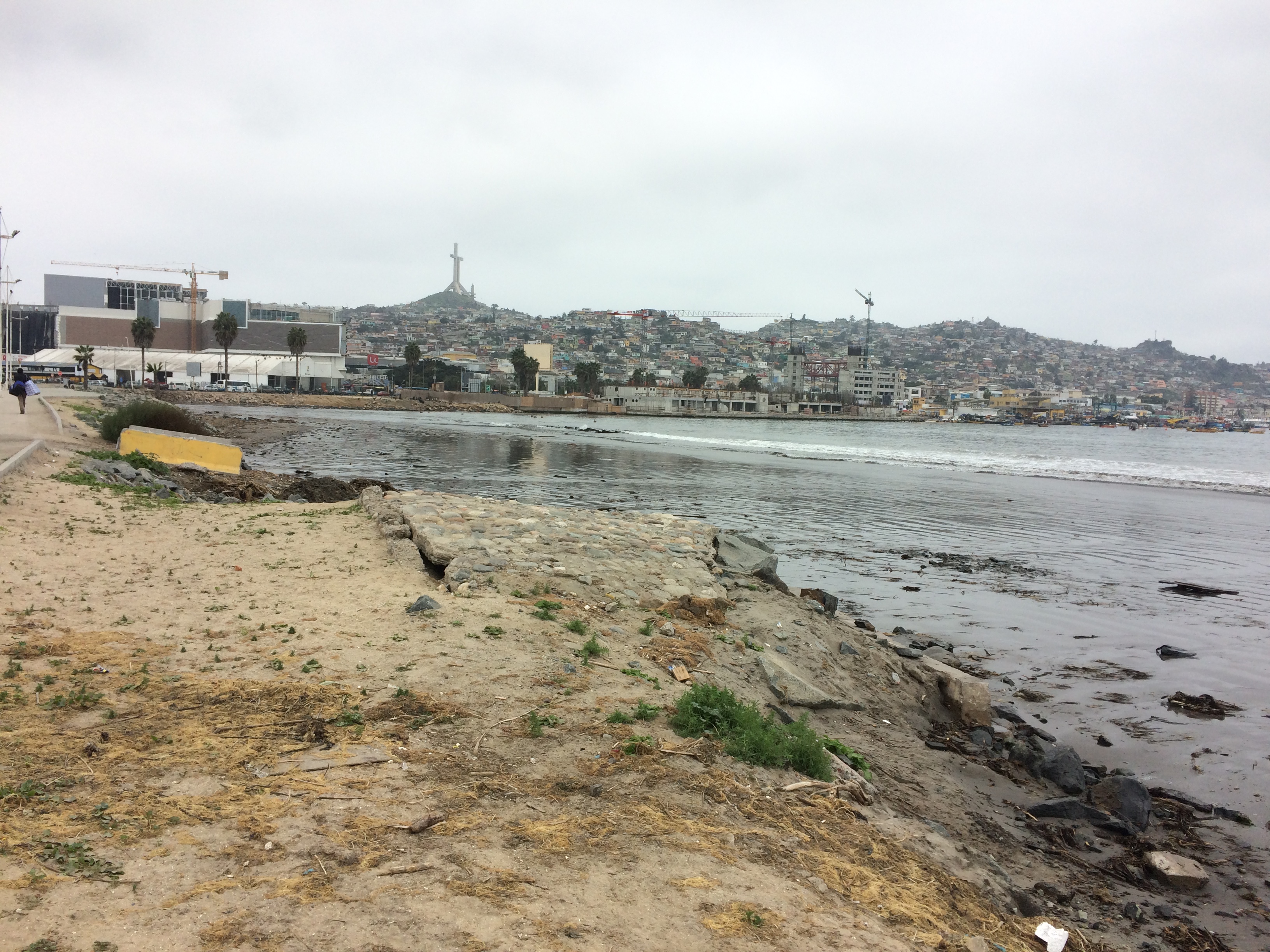}}
  	\subfloat[]{
  		\includegraphics[width=0.5\textwidth]{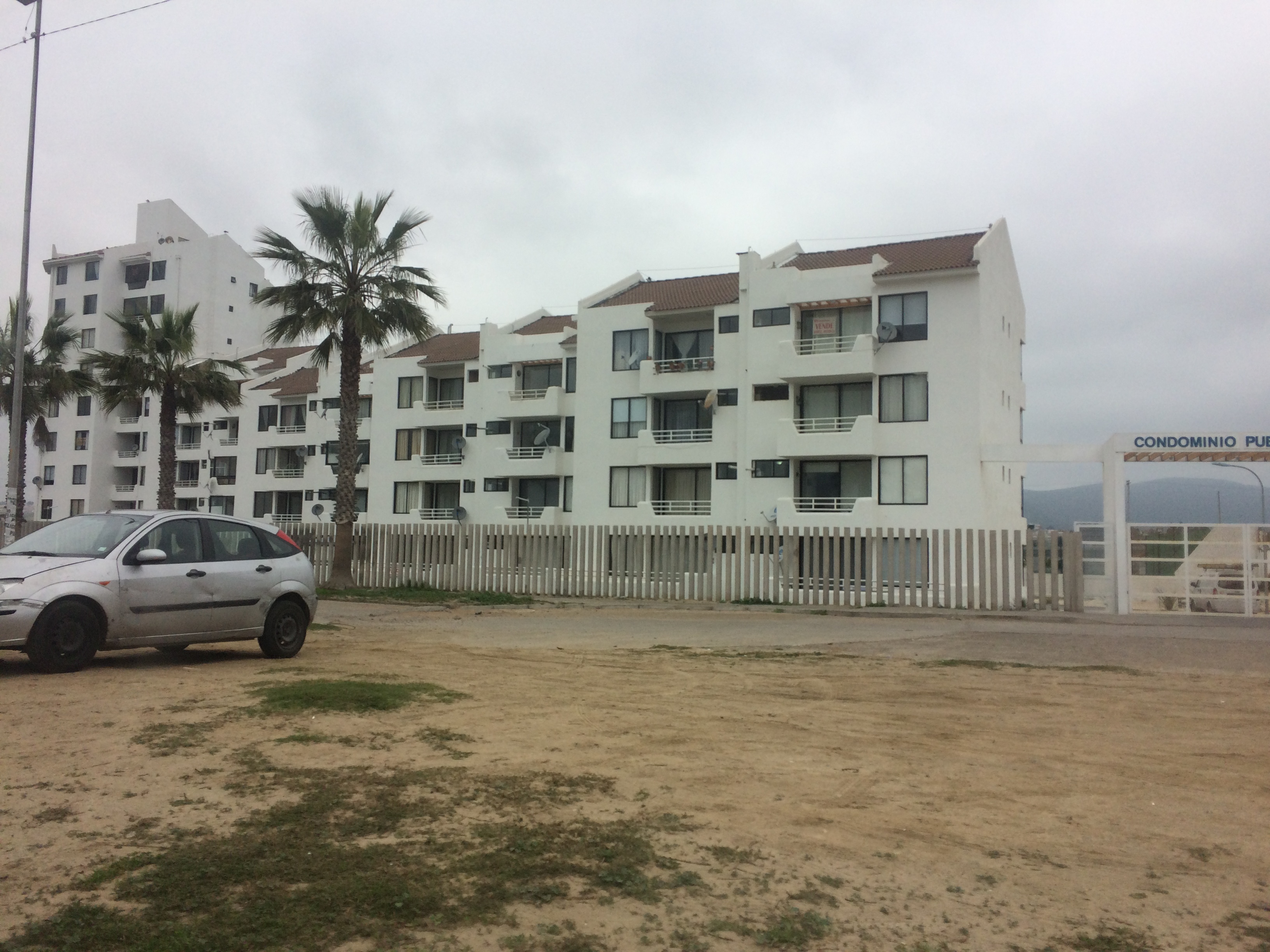}}\\
  	
  	\caption{Areas of the cities of La Serena and Coquimbo affected by the 2015 tsunami.}
  	\label{fig:tsu1922}
  \end{figure*}

 Figure \ref{fig:tsu1922} shows some areas that have been affected by the 2015 tsunami in the cities of Coquimbo and La Serena. Figure \ref{fig:tsu1922}.a shows one of the buildings that recorded an inundation depth of 1.80 m; Figures \ref{fig:tsu1922}.b and \ref{fig:tsu1922}.c shows zones that were affected in the backyard of the market and the beach, respectively; Figure \ref{fig:tsu1922}.d shows some buildings located along the beach, which do not have any protective structure.
 
  \section{Methodology}	
  
  To calculate the hydrodynamic tsunami force on a building using SPH simulations, it was necessary to calibrate the SPH model using hydrodynamic variables, such as the tsunami amplitude, flow velocity and flow direction obtained from a validated reference model. Based on the field data of the 2015 tsunami, the reference model allowed to obtain hydrodynamic variables in areas where no data was recorded. Subsequently, with the SPH model and using high resolution in topography and morphology of the beach, a more precise hydrodynamic behavior of the tsunami was obtained that better represented the flow interaction with the building.
  
   \subsection{Simulation Domain of the SPH model}
   
   Before defining the simulation domain of the SPH model, flow directions were evaluated at points located in the area of interest, based on the reference model simulation.
   This allowed to obtain an average flow direction of the tsunami waves from N $ 28 $ E with a minimum standard deviation in the area of interest during the tsunami, and to redefine new monitoring points aligned with the flow direction.
   
\begin{figure}
   	\centering
   	\includegraphics[width=0.8\textwidth]{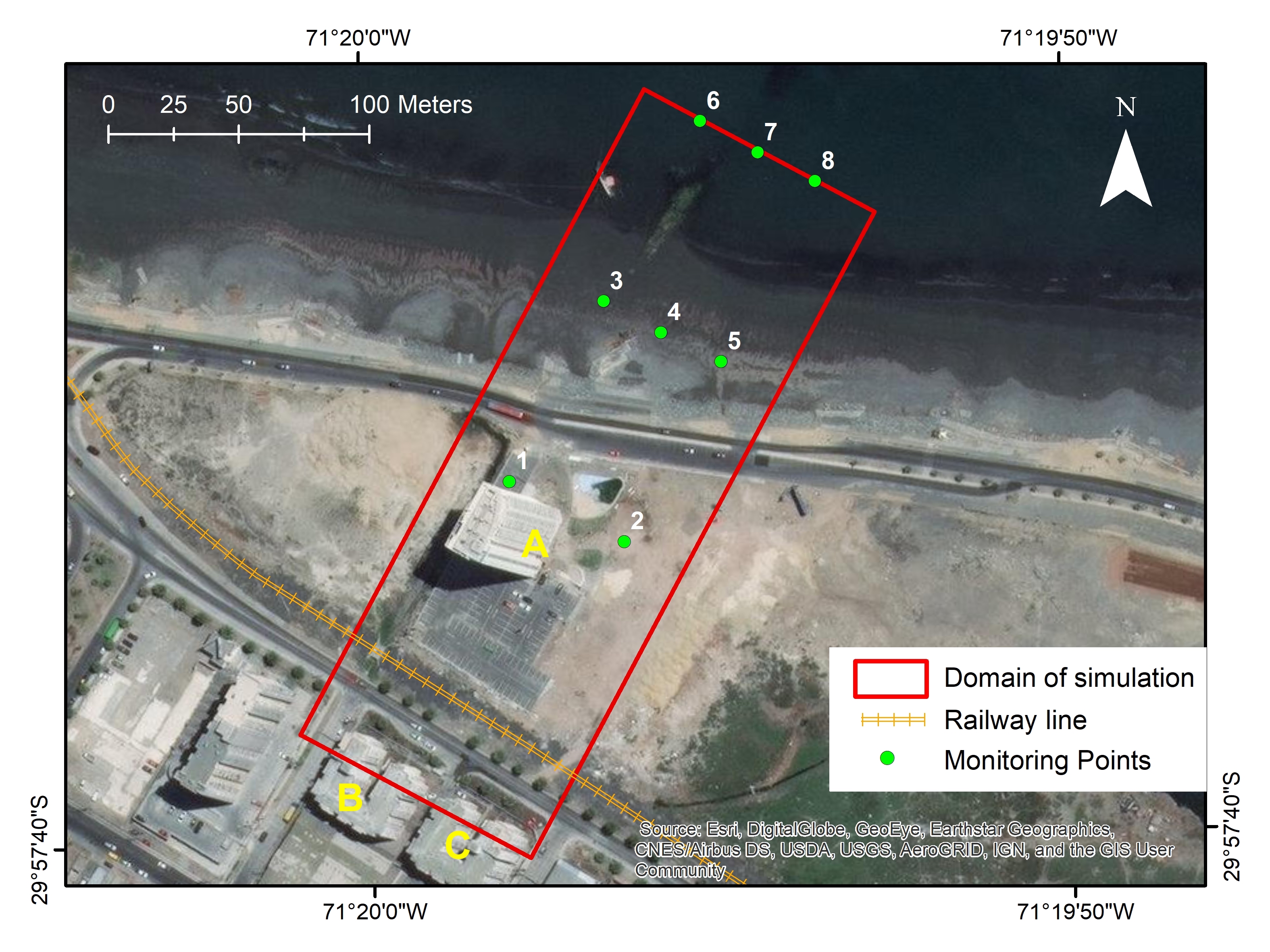}
   	\caption{Simulation domain of the SPH model; Monitoring points: M1, M2, M3, M4, M5, M6, M7, M8, M9 and M10; and three buildings for monitoring forces A, B and C.}
   	\label{fig:dominio1}	 
   \end{figure}

   Subsequently, the domain was defined with dimensions of $ 50 $ m by $ 300 $ m (Figure \ref{fig:dominio1}). This domain includes the beach, the railway and three buildings, and eight monitoring points for hydrodynamic variables.
   Points M1, M2, M3, M4 and M5 were considered for the validation and verification of the results of the SPH model with the reference model; whereas the points M6, M7 and M8 were considered to obtain the flow inlet conditions, from the reference model, into the domain of the SPH simulation.

   \subsection[Reference information]{Reference information for the 2015 tsunami in Coquimbo}
   
   The reference model corresponds to the numerical simulation carried out by Aranguiz et al. (2017). They used the Non-hydrostatic Evolution of Ocean WAVEs model (NEOWAVE) (Yamazaki et al., 2011; Yamazaki et al., 2009). This model solves the nonlinear shallow water equations and considers a vertical velocity term to account for weakly dispersive waves. The numerical simulation was validated by means of tsunami waveforms at Coquimbo and Valparaiso tide gauges, as well as with the DART buoy 32402 data. Digital elevation models, which define the domain morphology to be simulated, were created from bathymetry of nautical charts and topography from a detailed Digital Terrain Model from LIDAR data provided by the Coquimbo office of the Ministry Housing (MINVU). The discretization of the domain was done using 5 nested grids of different resolutions, ranging from 3600 m (in the Pacific Ocean) to 10 m (for the topography of Coquimbo). The roughness coefficient was defined as $n=0.025$ at both the sea bottom and urban area, according to a sensitivity test of the roughness coefficient, validated with tsunami inundation height results (Aranguiz et al. 2017).
   \\\\
   \textbf{Results of the reference model}\\	
   The 2015 tsunami was characterized by a series of large waves, however, not all of them approached the coast far inside. Through the reference model, five waves were identified that caused large flows in one of the topographically vulnerable areas, 100 meters away from the beach. These waves occurred approximately every 35 minutes, and affected the coast during several hours.
   
\begin{figure}
    	\centering
    	\includegraphics[width=0.8\textwidth]{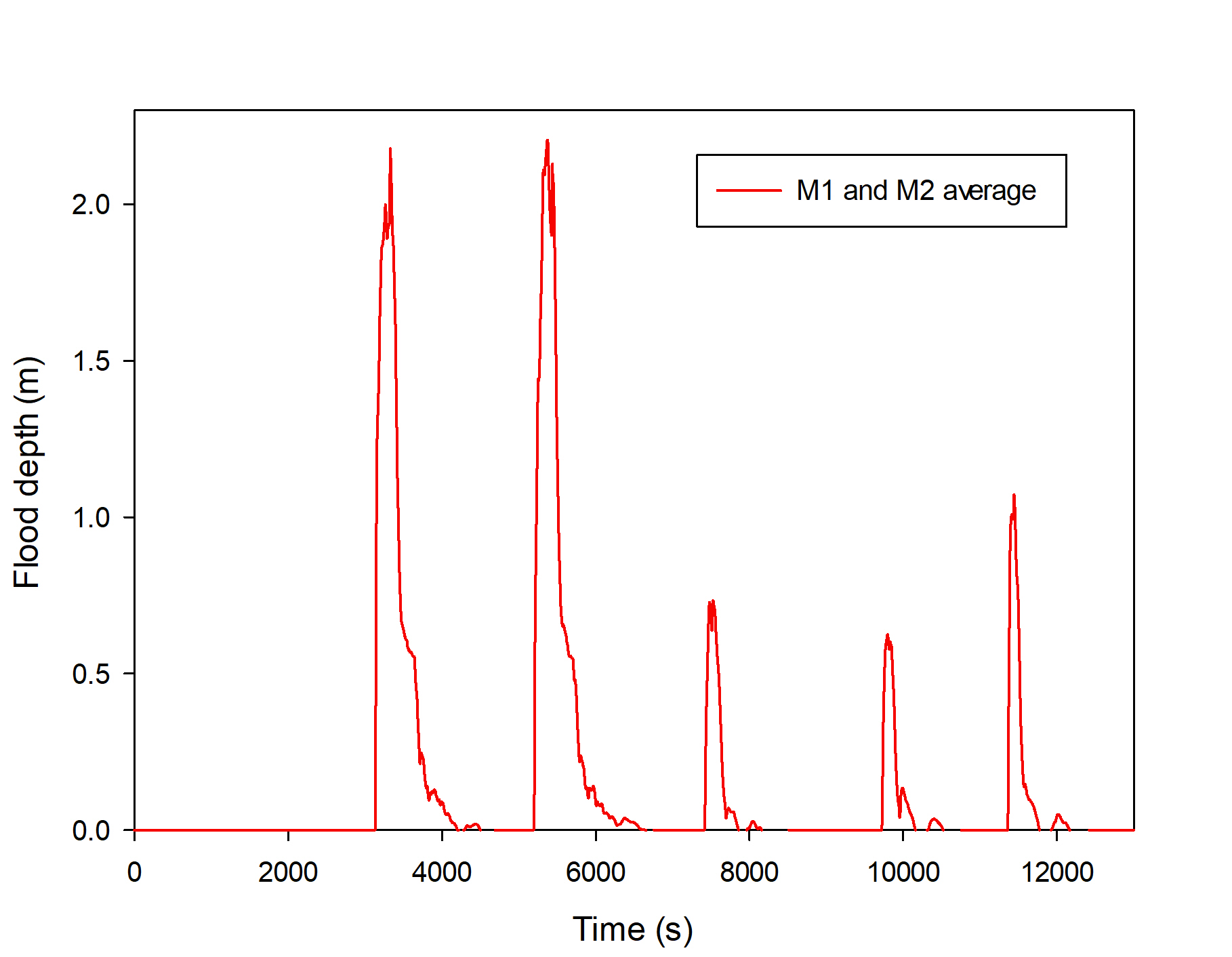}
    	\caption{Temporal evolution of the flood depth in two points located on the beach (at the ground level). }
    	\label{fig:tewd}	 
    \end{figure}

   In order to calibrate the SPH model simulations, flood depth and flow velocity values calculated by the reference model simulations at several points around the building were considered. Figure \ref{fig:tewd} shows the reference model results for the temporal evolution of flood depth, average of two points located on the beach (M1 and M2, Figure \ref{fig:dominio1}), near building A. The first and second waves have similar behaviour, nevertheless the second wave produced the largest flood depths near the building (at the monitoring points) and it has a slightly longer duration than the first wave. The second wave was selected to be simulated with the SPH model.
   
\begin{figure}
    	\centering
    	\includegraphics[width=0.8\textwidth]
    	{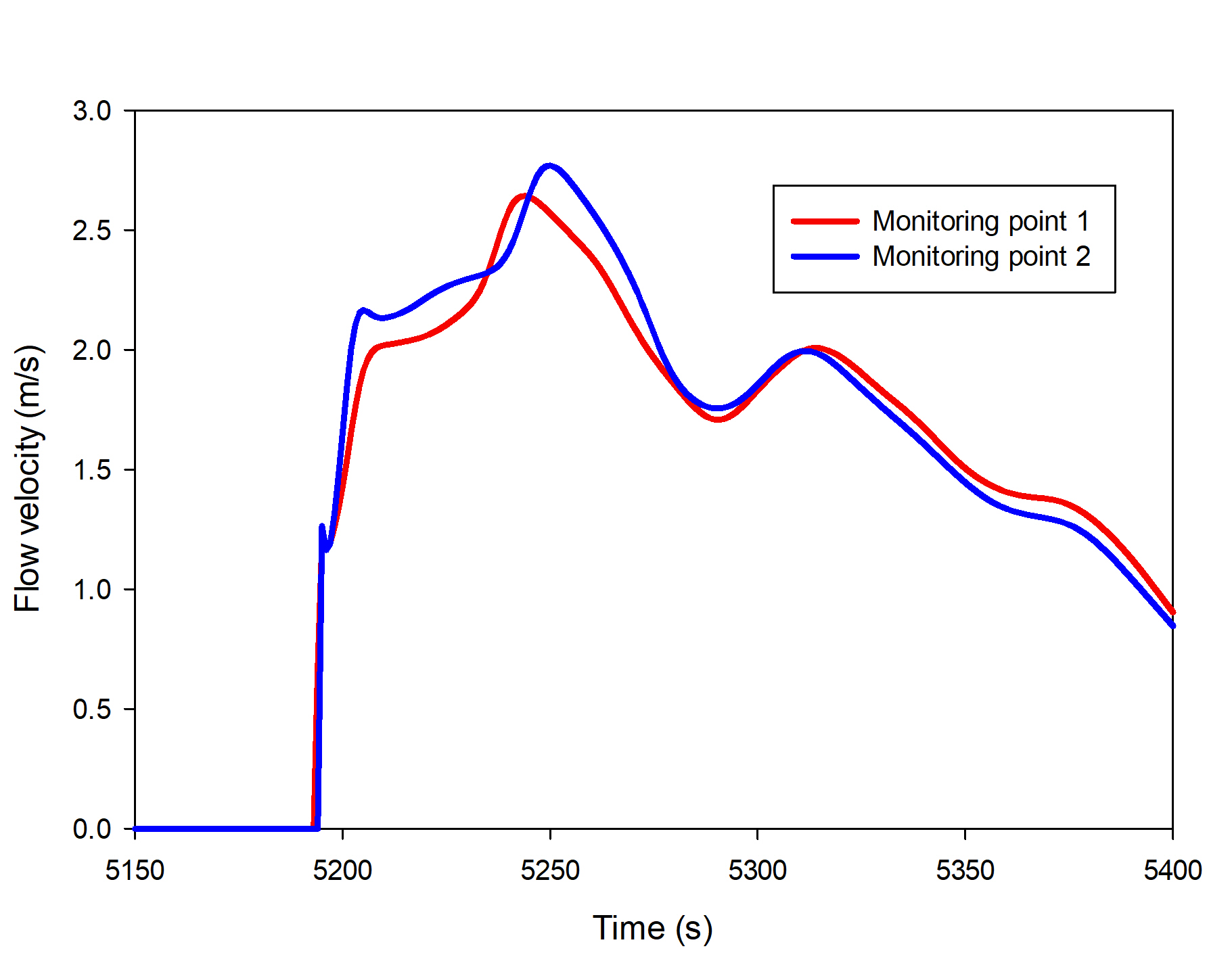}
    	\caption{Flow velocity in two points on the beach, reference model simulation data.}
    	\label{fig:vnsd}	 
    \end{figure}

    Figure \ref{fig:vnsd} shows the velocity time series at the monitoring points M1 and M2 during the second tsunami wave. The velocity difference between the two points is minimal, showing little spatial variation in the vicinity of the building under analysis.
    
    \subsection{Parameters and SPH simulation conditions}	
    
    In order to optimize the SPH model simulations, different interparticle spacing tests were performed varying the size of the particles of the SPH model
    and an acceptable approximation for the hydrodynamic variables was obtained with a interparticle spacing of 0.3 m. This resolution has been also used by Gonzalez-Cao et al. (2018) in 1:1 scale simulations.

    In this work, three types of SPH model simulations of the 2015 tsunami in Coquimbo were carried on:
    
    \begin{description}
    	\item[SPH model a:] this simulation considers the same topography of the reference model, i.e, without considering the details of the railway line and buildings. The objective of this simulation is to corroborate the results of both models (flood depth and flow velocity at points M1 and M2).
    	\item[SPH model b:] this simulation includes the topographic detail of the first model as well as the presence of three buildings. 
    	\item[SPH model c:]  this simulation considers the presence of buildings used in the previous simulation with a greater topographic resolution such that the railway is well represented. 
    \end{description}
    
    \textbf{Input conditions}
    
    The tsunami wave in Coquimbo (Figure \ref{fig:tewd}) was generated by means of a piston located at the northern boundary of the domain. When the piston moves vertically, it generates a volume of water on the level of the sea. 
    The movement of the piston was set to be vertical in order to optimize the number of contour particles and reduce the computational time.

    According to the velocity of the piston movement, different flow velocities and flood depths are generated at the reference points. Therefore, the piston velocity was defined in such a way that the hydrodynamic variables of the SPH model are in good agreement with the reference model variables. The long period wave that represents the second tsunami wave of the 2015 event in Coquimbo was defined by means of a flow function obtained from flow velocity and flood depth data computed at the points $6$, $7$ and $8$.
    It should be mentioned that the input flow is considered in the domain at the points 6, 7 and 8.
    
\begin{figure}
    	\centering
    	\includegraphics[width=0.9\textwidth]{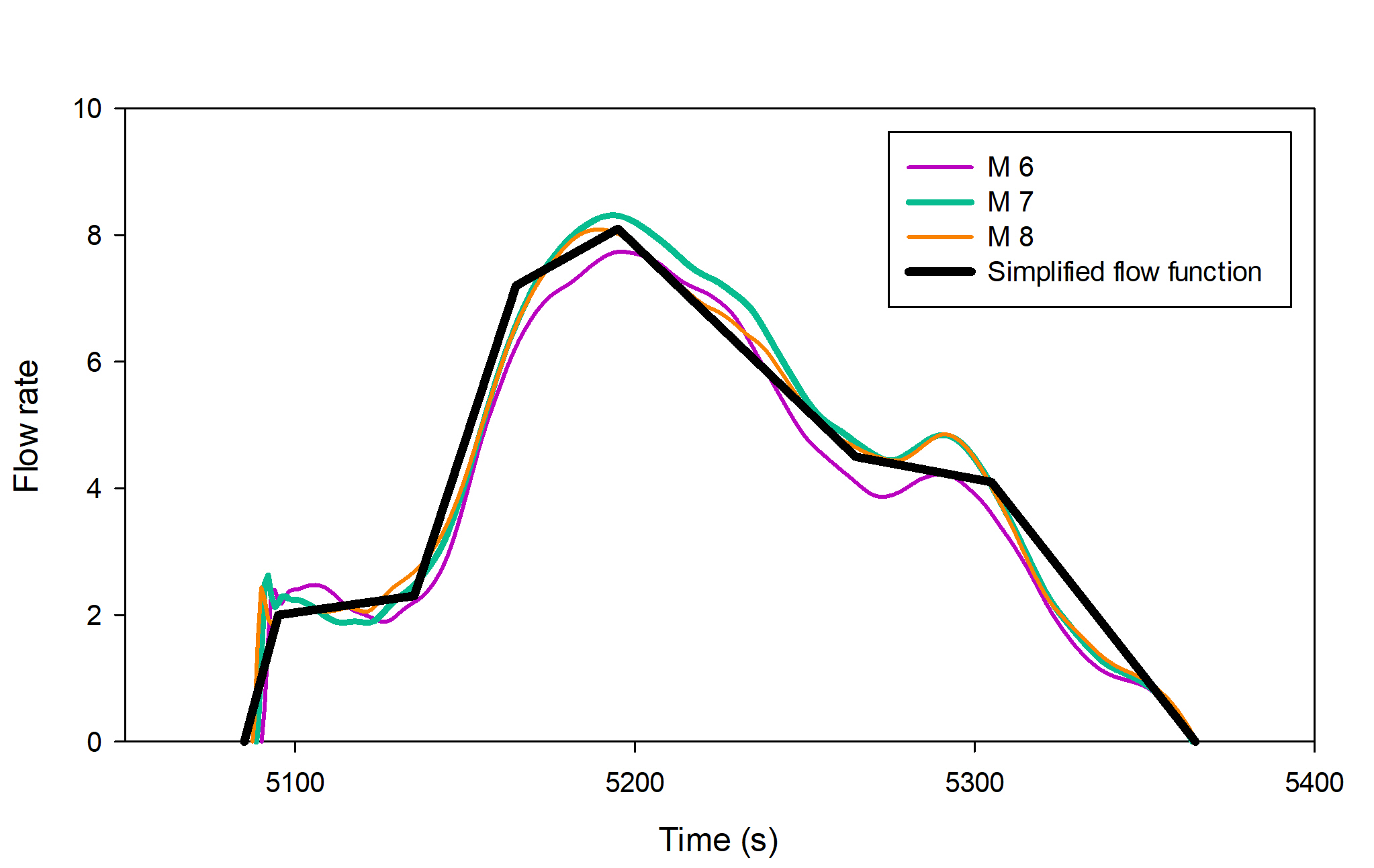}
    	\caption{Flow Rate Functions}
    	\label{fig:cases}	 
    \end{figure}

\begin{figure}
    	\centering
    	\includegraphics[width=0.8\textwidth]{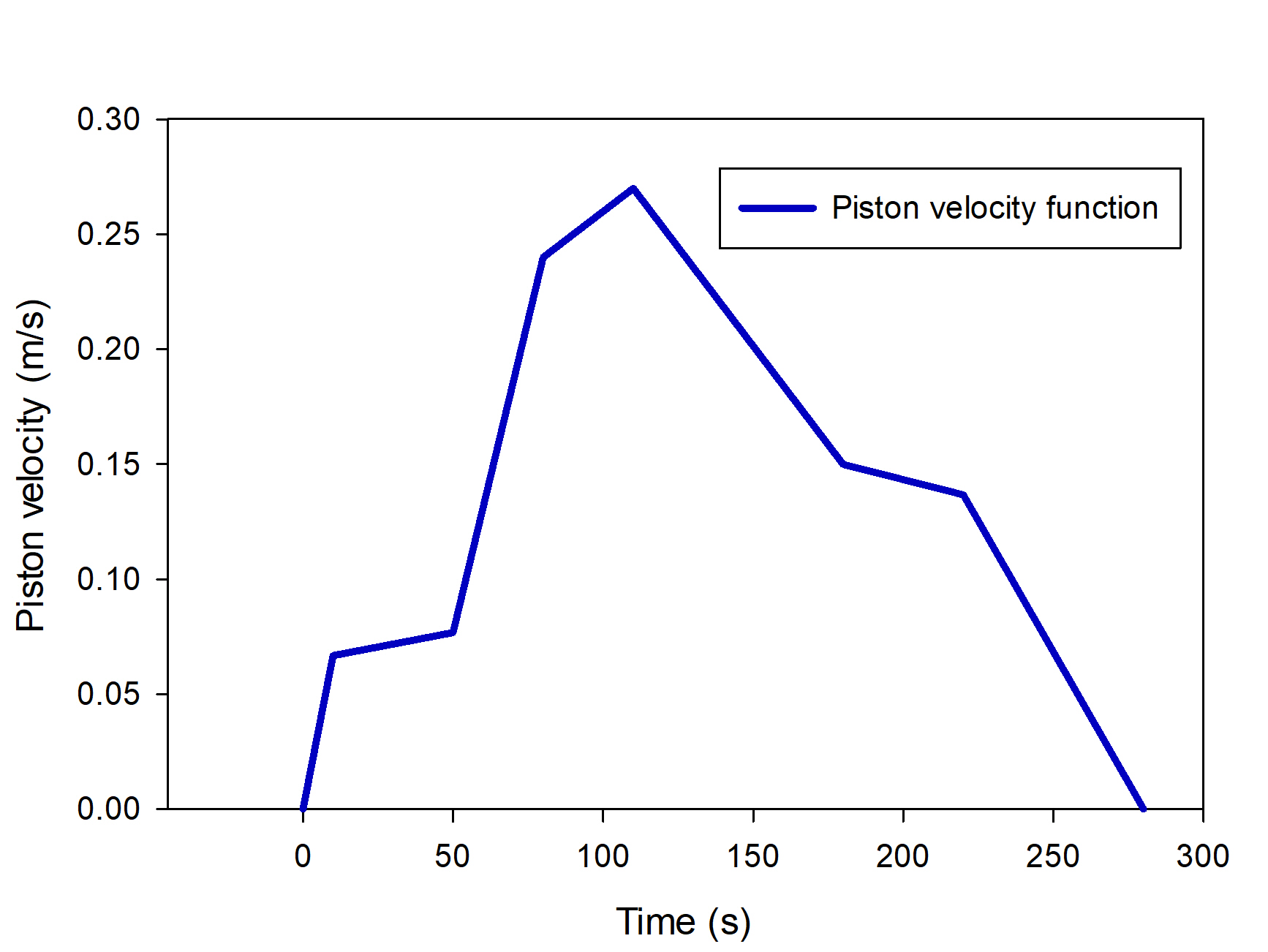}
    	\caption{Piston Velocity Function.}
    	\label{fig:vel_pis}	 
    \end{figure}

    Figure \ref{fig:cases} shows the flow variation given by the reference model that enters the domain at points $6$, $7$ and $8$. The black line represents a simplified input flow function that needs to be reproduced by the piston in the SPH model to obtain similar hydrodynamic variables of the tsunami. The velocity and acceleration of the piston are chosen to reproduce the simplified flow function. Figure \ref{fig:vel_pis} shows the final piston velocity function used in the SPH simulations. The maximum unit flow presented in Figure \ref{fig:cases} (t= $5200 s$) is $8m^2/s$, since this yields a piston maximum velocity  
$(8m^2/s) / (30m) = 0.267 m/s$, as depicted in 
Figure \ref{fig:vel_pis}).

    As an initial condition the flood height was set to sea level with respect to the topography information.
    
    In the DualSPHysics code, boundary conditions that describe the fluid boundary are considered as fixed particles that limit the fluid particles.
    Both the boundary of the domain and the walls of the buildings were constructed using fixed particles.
    
    As an output condition, the south zone of the domain is left as an open contour. This allows the fluid particles to escape when they are close because of the flow.
    In the south domain two buildings were introduced, which slow down the flow and increase the flood elevation.
    
    The parameters and characteristics of the simulation are shown in Table \ref{parametros_des}. 
    
    \begin{table}
    	\centering
    	\caption{Parameters and main characteristics of the SPH Model.}
    	\begin{tabular}{lll}
    		\hline
    		Parameters  & Value \\
    		\hline 
    		Kernel Function & Quintic\\ 
    		Time-step Algorithm & Verlet \\ 
    		Viscosity & Artificial  $\alpha=$0.01\\ 
    		Fluid particle size & 0.3m \\ 
    		Number of particles º & 2657089 \\ 
    		Simulated time ª & 300s \\ 			
    		Computing time & 5 days \\ \cline{1-2}							
    	\end{tabular}
    	\label{parametros_des}
    \end{table}		 
    
    \subsection[Validation]{Validation of the SPH model with respect to the reference model}
\begin{figure*}
    	\centering
    	\subfloat[]{
    		\includegraphics[width=0.9\textwidth]{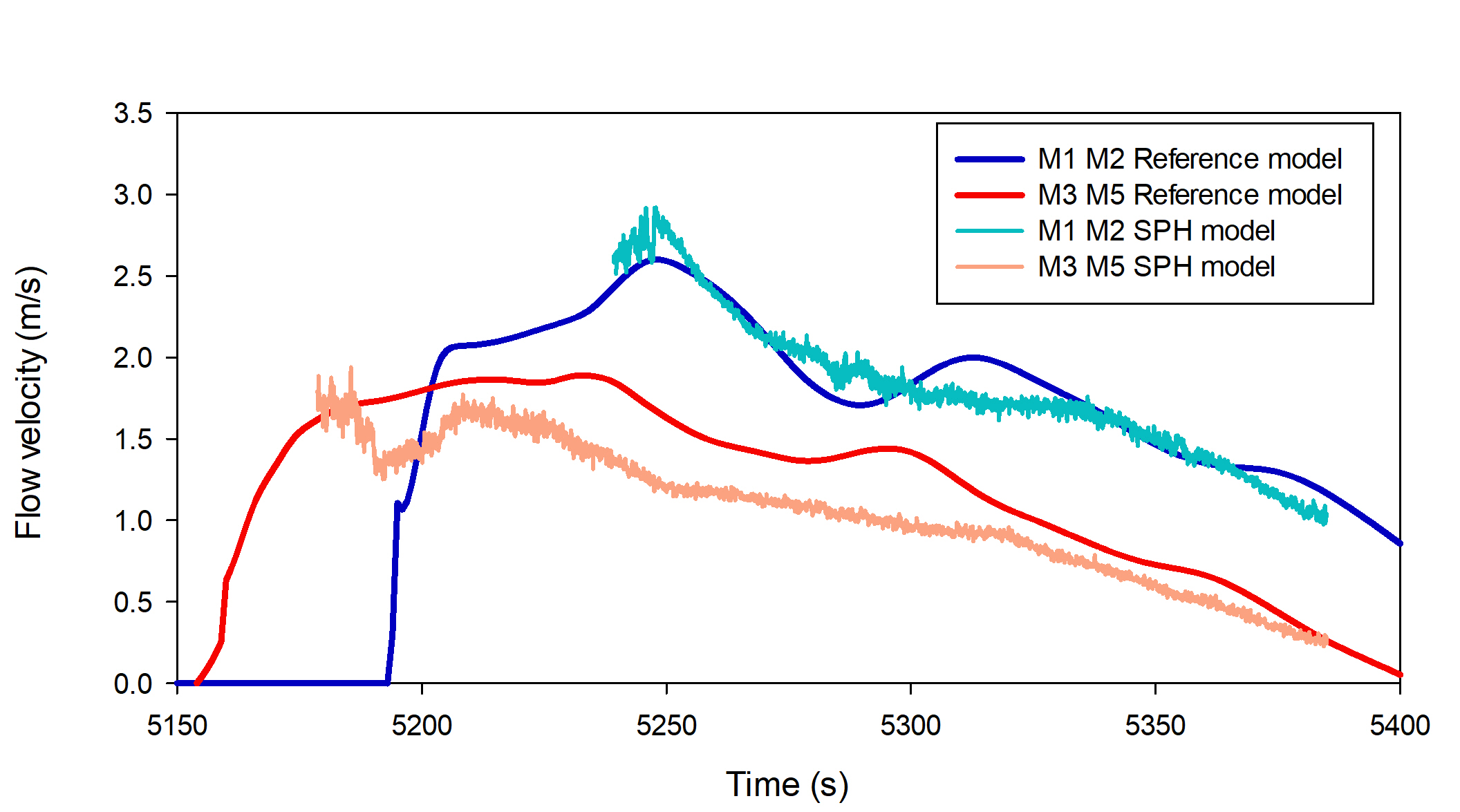}}\\
    	\subfloat[]{
    		\includegraphics[width=0.9\textwidth]{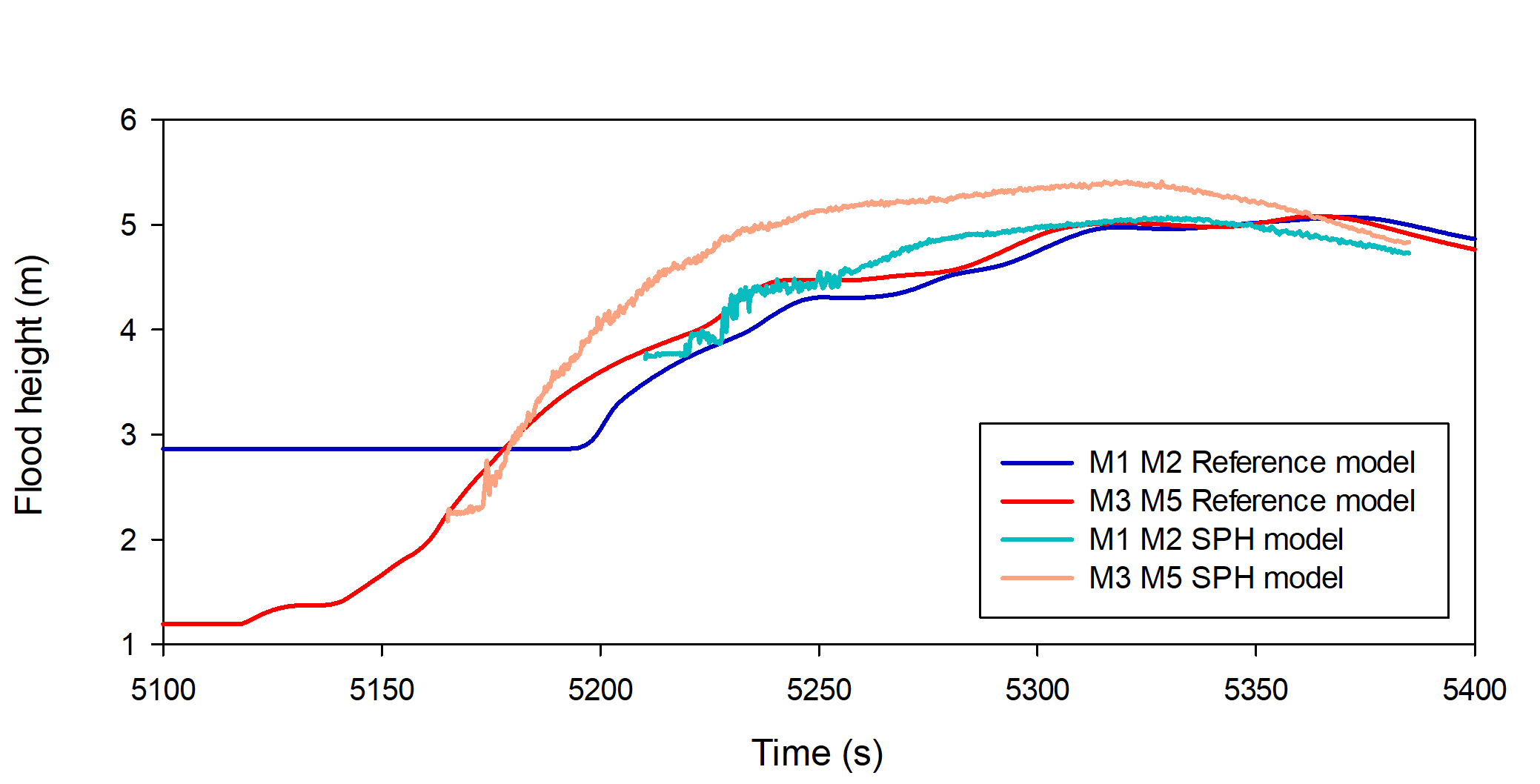}}
    	\caption{Comparison of the SPH model case a and the reference model results. a$) $ Flow velocity at points M1-M2 and M3-M5. b$) $ Flood depth at points M1-M2 and M3-M5.}
    	\label{fig:compar1}
    \end{figure*}

    The comparison between the SPH model and the reference model flow velocity and flood depth is shown in Figure \ref{fig:compar1}, considering the same topographical conditions and without buildings on the SPH model a. The comparison show results in two zones, the first at 40 meters from the beach (Points M1 and M2, Figure \ref{fig:dominio1}) and the second zone at 120 meters from the beach (points M3 and M5, Figure \ref{fig:dominio1}).
    Results for points M1 and M2 are grouped together, as are points M3 and M5, since their topographic elevation, their distance to the beach and consequently their hydrodynamic values are similar. In order to compare results from both models, the flow velocity of the SPH model is averaged in depth.
    
    Results indicate that for the points M3 and M5, the flow velocity in the SPH model simulation is lower than the velocity obtained in the reference model; however the flood depth is a bit higher in the SPH model, due to mass conservation. A better correspondence is shown, both in velocity and flood depth, for the points close to the building A (points M1 and M2). 
    
    The validation shows a good correspondence in flow velocities and flood depths between the SPH model and the reference model. However, there is a small discrepancy in the values, which are probably due to the simplification of the flow direction in the SPH model.
    
    \section{Results of SPH simulations b and c}	
    
    In this section the hydrodynamic forces on buildings are numerically estimated and compared with the Chilean standard ChN3363. 
    
    Once the numerical parameters necessary to achieve the hydrodynamic correspondence of the SPH model with the reference model were calibrated (see section 3.4), the geometry of the SPH model was modified adding three buildings (SPH model b) and, subsequently improving the topography considering the railway line (SPH model c), which works as a barrier to the flow (Figure \ref{fig:dominio1}).

    \begin{figure*}
    	\centering
    	\subfloat[]{
    		\includegraphics[width=0.9\textwidth]{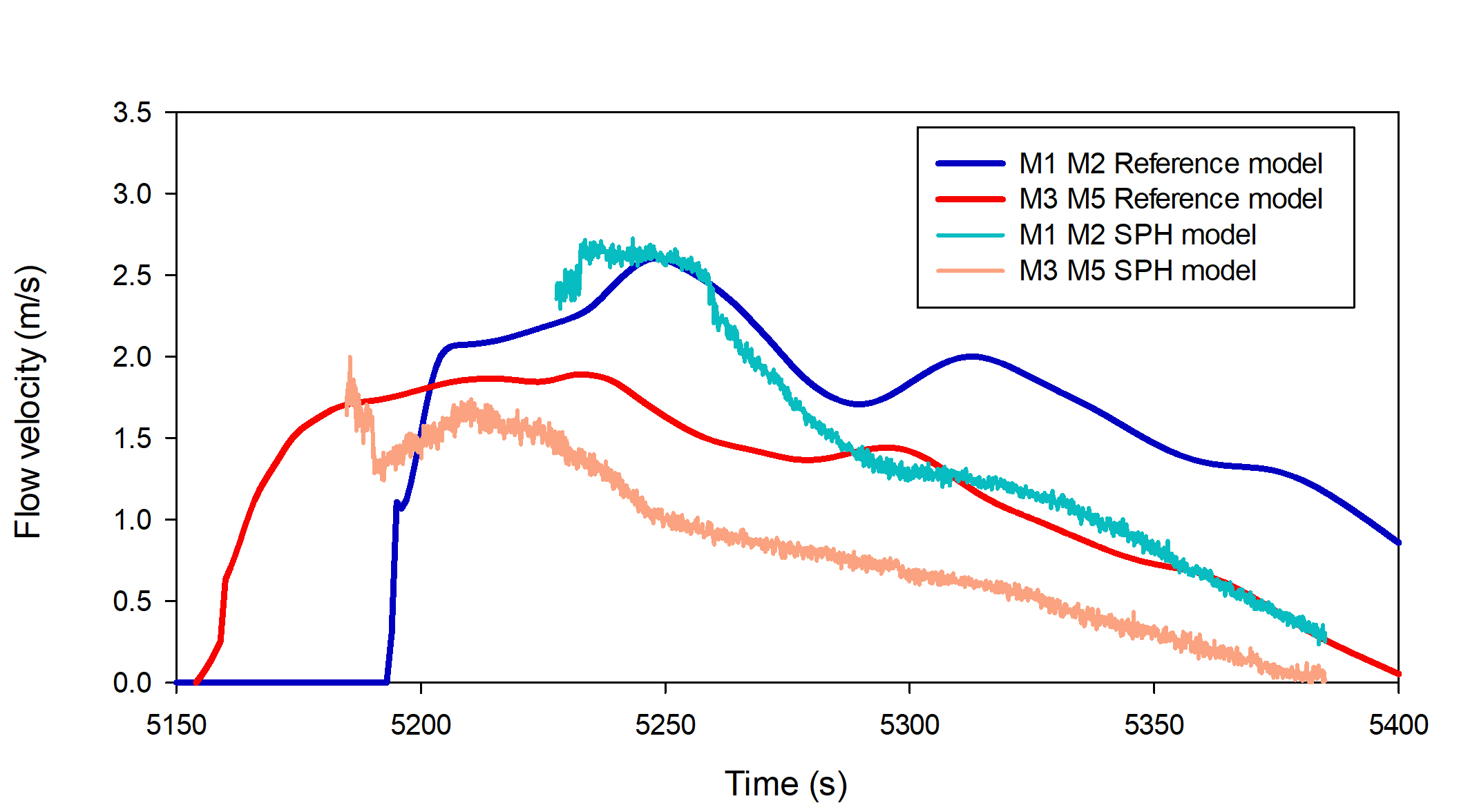}}\\
    	\subfloat[]{
    		\includegraphics[width=0.9\textwidth]{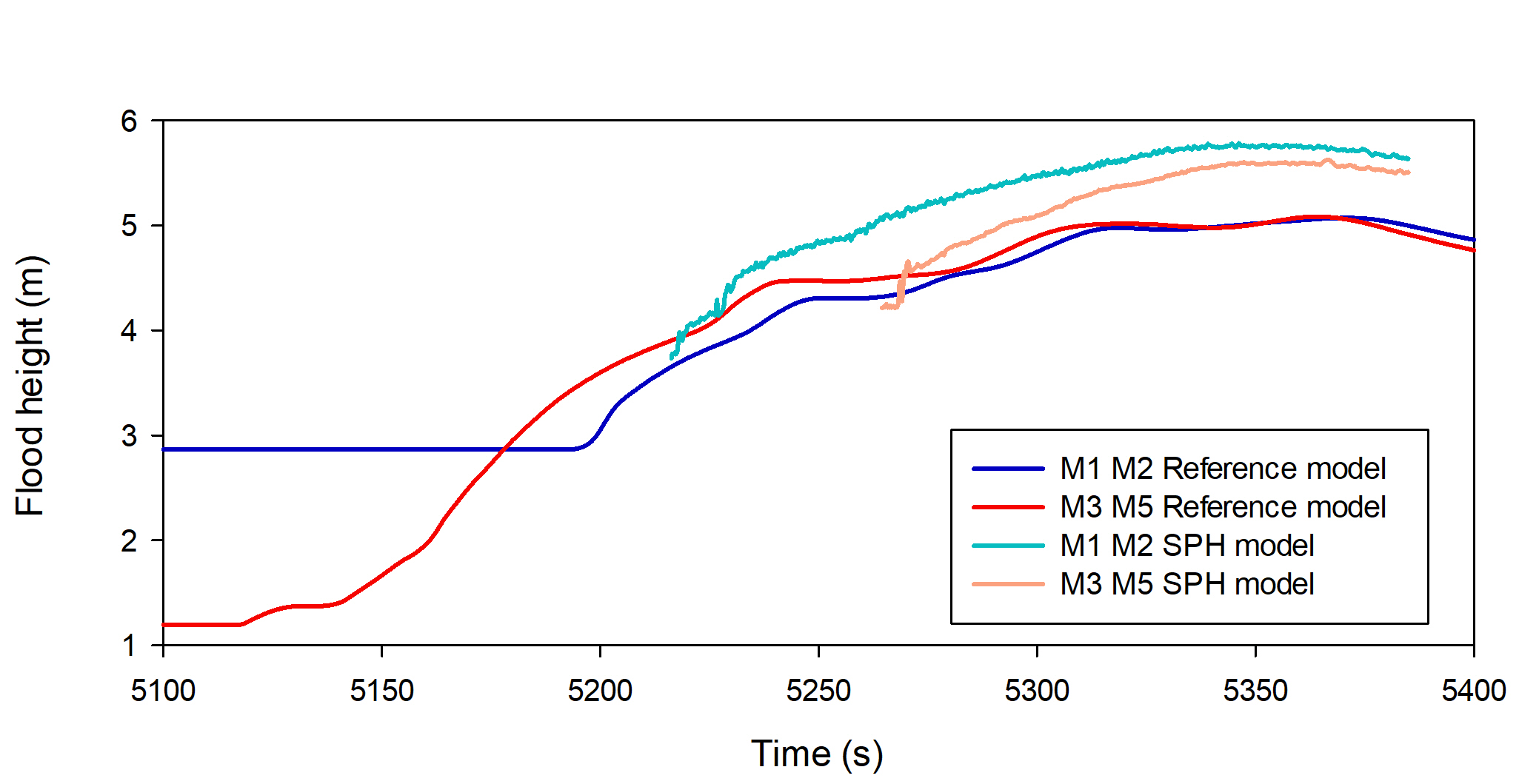}}
    	\caption{Comparison of the SPH model c and the reference model results. a$)$ Flow velocity at points M1-M2 and M3-M5. b$)$ Flood depth at points M1-M2 and M3-M5.}
    	\label{fig:compar2}
    \end{figure*}
    
\begin{figure}
      	\centering
      	\includegraphics[width=0.5\textwidth]{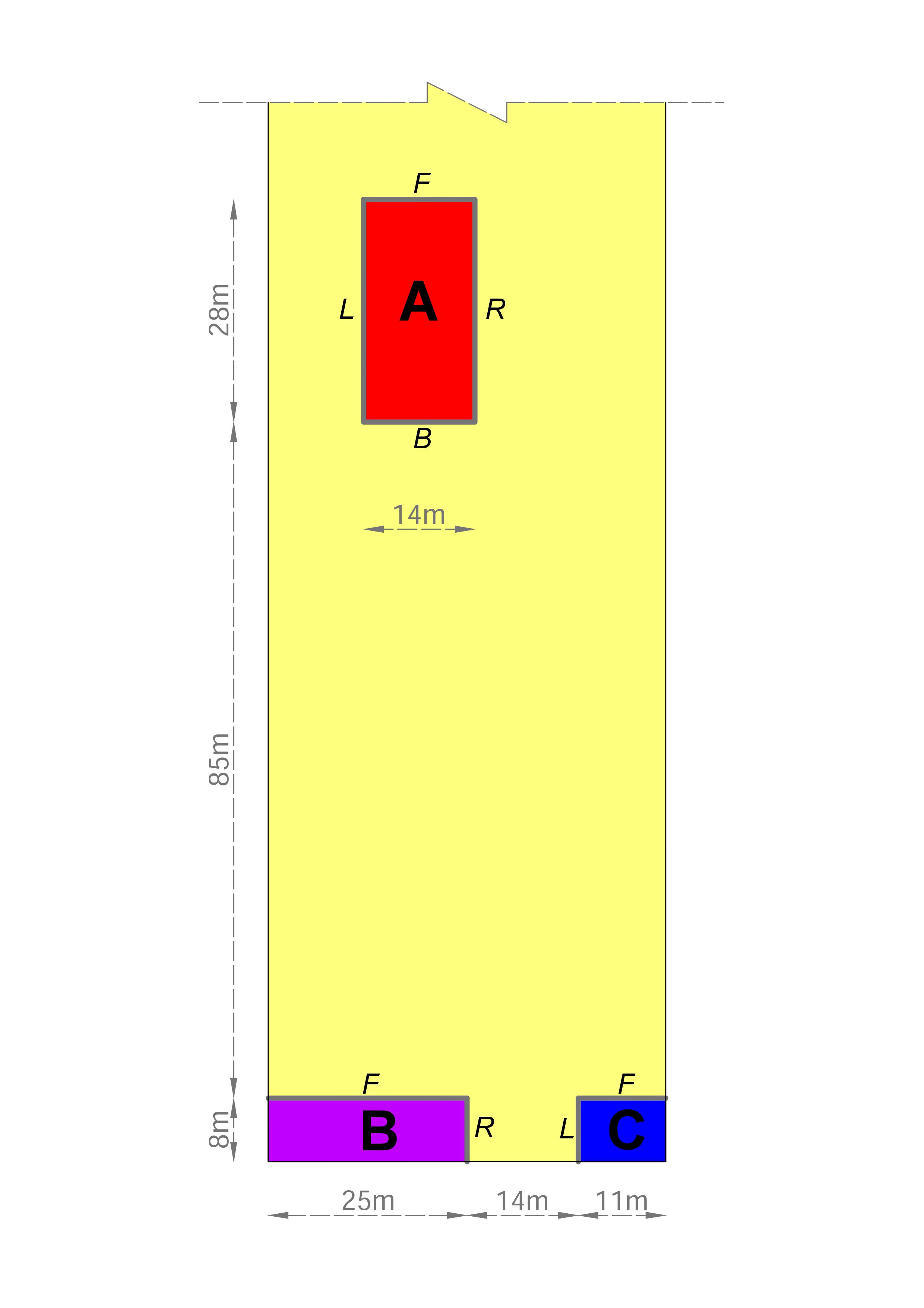}
      	\caption{SPH model scheme, buildings within the simulation domain.}
      	\label{fig:esquema}	 
      \end{figure}

    For the building A the forces excerted by the tsunami were calculated on the faces shown in Figure \ref{fig:esquema} (F: front, B: back, L: left and R: right.). 
    
       \subsection[Comparison]{Numerical and empirical comparison}
       
\begin{figure}
       	\centering
       	\includegraphics[width=0.8\textwidth]{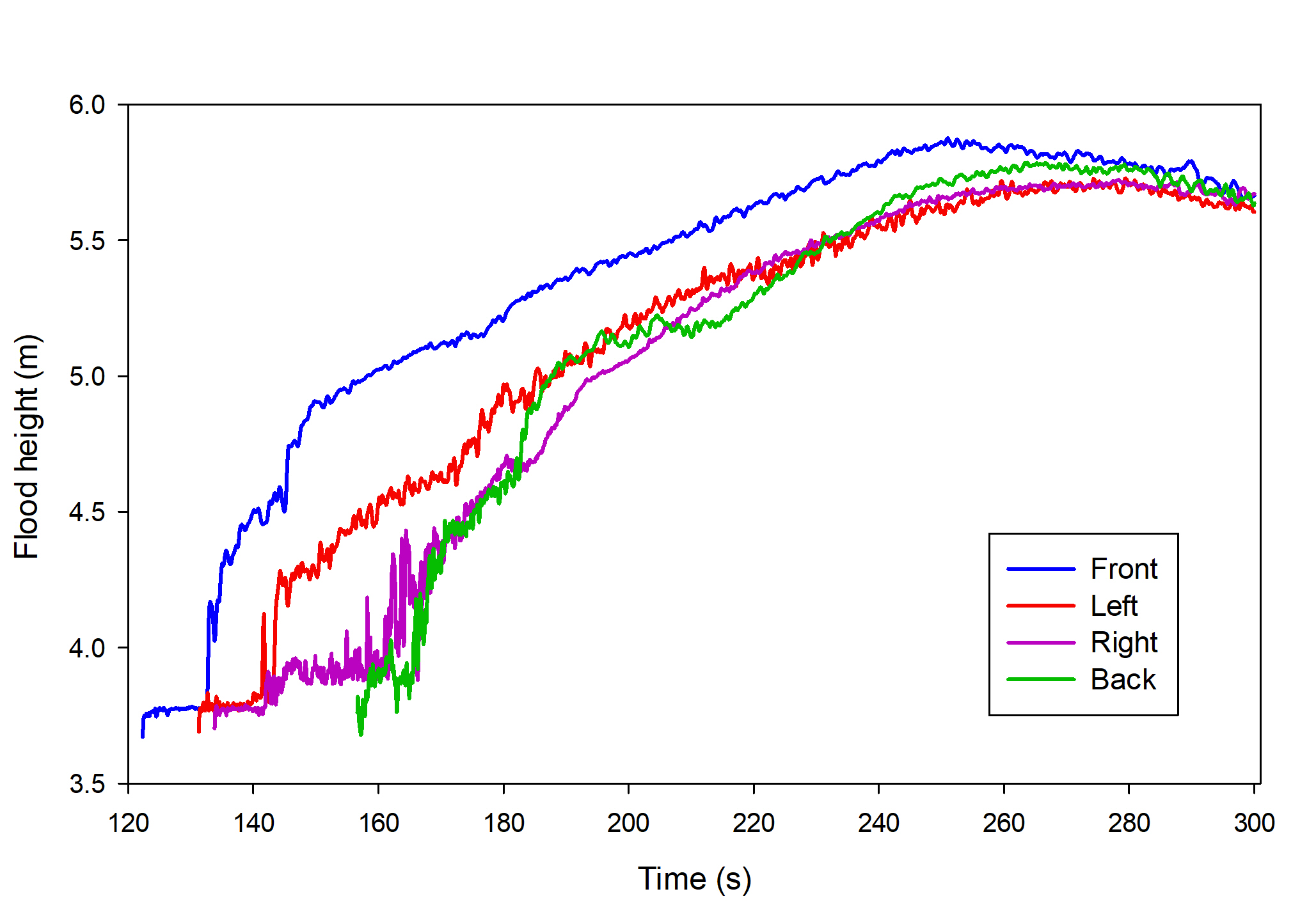}
       	\caption{Numerical flood heights next to building A, for the SPH model case c.}
       	\label{fig:nextA}	 
       \end{figure}

       The Figure \ref{fig:nextA} shows the flood height on building A for the SPH model c case, the front corresponds to the face of the building that receives the impact of the tsunami. The evolution of the wave produces maximum flood heights up to four minutes later from the instant that the water reaches building A.
       It is shown that in the front face the flood height increases faster, and at all moments it is greater than in the other faces of the building. Because of the small variations that exists on the sides of the building, the left side has slightly larger flood heights compared to the opposite side. Finally, the back face is where the wave arrives later; however, soon after, flood heights on this face are higher than floods on the lateral sides. This is due to the accumulation of water in the back of the building that occurs when the water collides with the railway line.
       
\begin{figure}
       	\centering
       	\includegraphics[width=0.9\textwidth]{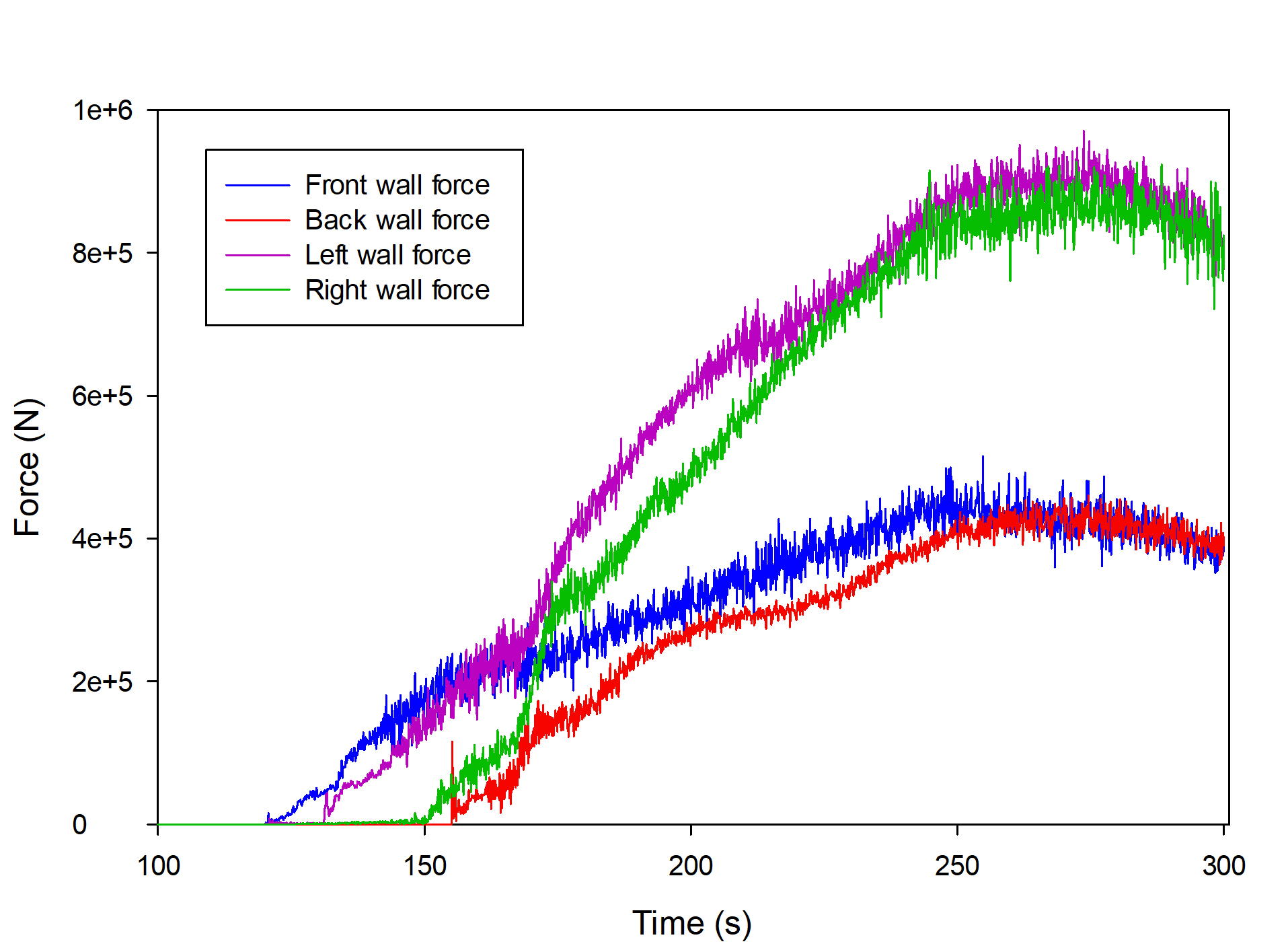}
       	\caption{Force on building A for the SPH model c.}
       	\label{fig:vel_FORCEa}	 
       \end{figure}

       SPH calculated forces on a building are obtained as the sum of forces acting on a surface (Eq. \ref{eq:dinF}). 
       The numerical results of the forces on building A, obtained from the SPH model c are shown in Figure \ref{fig:vel_FORCEa}. From the figure, it is possible to observe that the forces on the side walls of the building are generally higher than the the forces received by the front and back walls; however, it must be emphasized that this force corresponds to the integral of the force over the surface (see Eq. \ref{eq:dinF}) and this is due to the fact that the length of the side walls are 2.5 times the length of the front and back walls.	\\

\begin{figure*}
       	\centering
       	\includegraphics[width=1\textwidth]{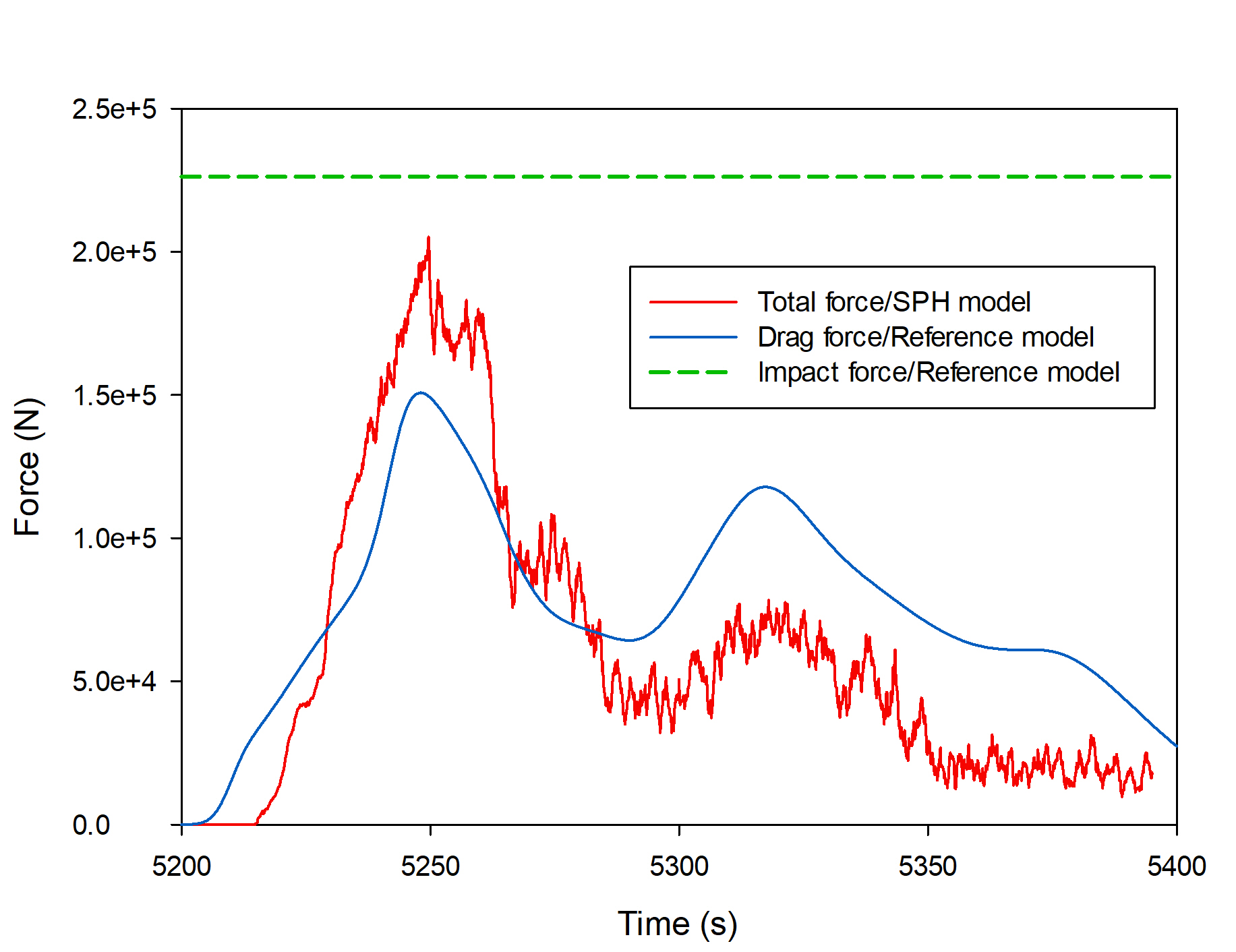}
       	\caption{Force estimation on the building A; forces from SPH: direct estimation by SPH model; drag force ($F_D$): empirical estimation of drag force using the output data of the reference model; impact force: empirical estimation of wave front force ($1.5*$DragForce, $F_I$).}
       	\label{fig:comp_for}	 
       \end{figure*}

       In order to compare numerically calculated hydrodynamic forces acting on building A with empirical values, the forces are estimated using the Chilean standard ChN3336.
       The Chilean standard estimate is composed by a set of equations that represent forces of different nature.
       The drag force, caused by the flow velocity around the structure, is calculated as $F_D=1/2 \ \gamma/g \ C_D \ \left(d \ v^2\right)$,
       where $C_D=2$ is a drag coefficient, $g$ is the gravity acceleration, and $v$ the flow velocity (averaged from velocities at points M1 and M2).
       \\
       The impact force is caused by the impingement of the leading edge of a tsunami wave affecting a structure. The impact force acts only on the front side of the structure. This force is computed as a function of the maximum drag force as $ F_S= 1.5 F_D$.
       \\\\
       Figure \ref{fig:comp_for} shows the comparison of empirical and SPH numerical estimations of hydrodynamic forces on building A. SPH maximum values of forces are higher than the drag forces calculated with the reference model. It can be explained considering that the reference model does not include the building in the simulation and, in addition, impact force is not considered in the drag force.
       Nevertheless, the hydrodynamic force of the SPH model does not become greater than the impact force estimated by the reference model (1.5 times the maximum drag force), this may be due to the fact that the topography and the buildings produce a decrease in the flow velocity (which is directly proportional to the drag force). 
       However, the impact force of the reference model is just above the results of the SPH model. Comparing the results of the drag force, the SPH model estimation is higher at the initial impact time of the flow in the building, but then it decreases below the drag force of the reference model, because the flow velocity in the SPH model decreases faster than in the reference model (due to the collision of the flow with the railway line).
       
       \subsection[Sensitivity of forces]{Sensitivity of the forces to the topographic resolution}
       In order to  show the sensitivity in the results of the SPH model produced by considering or not the railway line, forces acting on each wall of building for models b and c are compared.
       This allows estimating the variation of the hydrodynamic forces and determining if topographic simplification could underestimate or overestimate the forces over buildings.
       
       Once the detail of the railway line and the buildings are added to the topography of the SPH model c, a decrease in the flow velocity and an increase in the flood depth can be observed. This is due to the collision of the flow with the imposed objects. 
       
\begin{figure*}
       	\centering
       	\subfloat[]{
       		\includegraphics[width=0.5\textwidth]{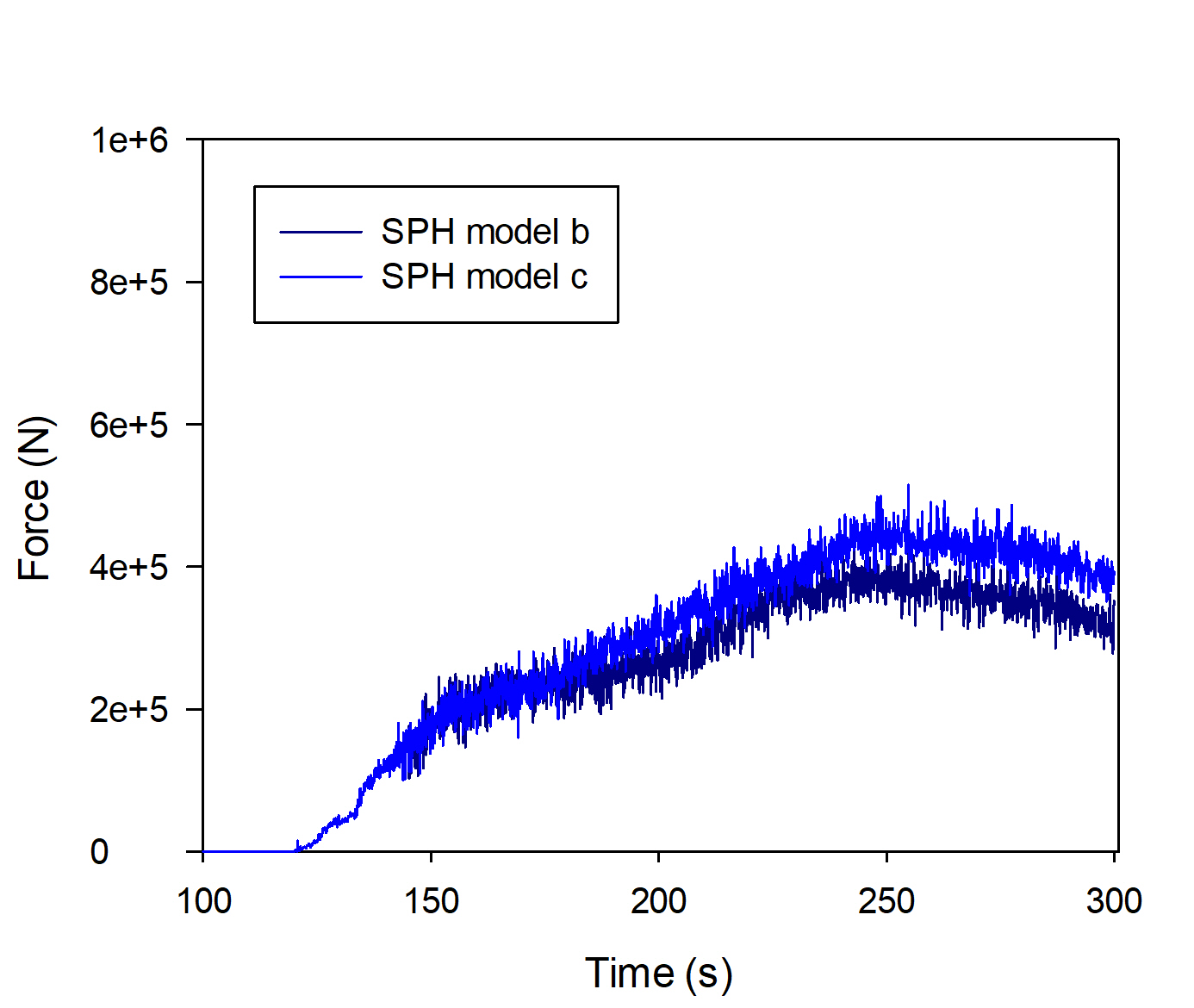}}
       	\subfloat[]{
       		\includegraphics[width=0.5\textwidth]{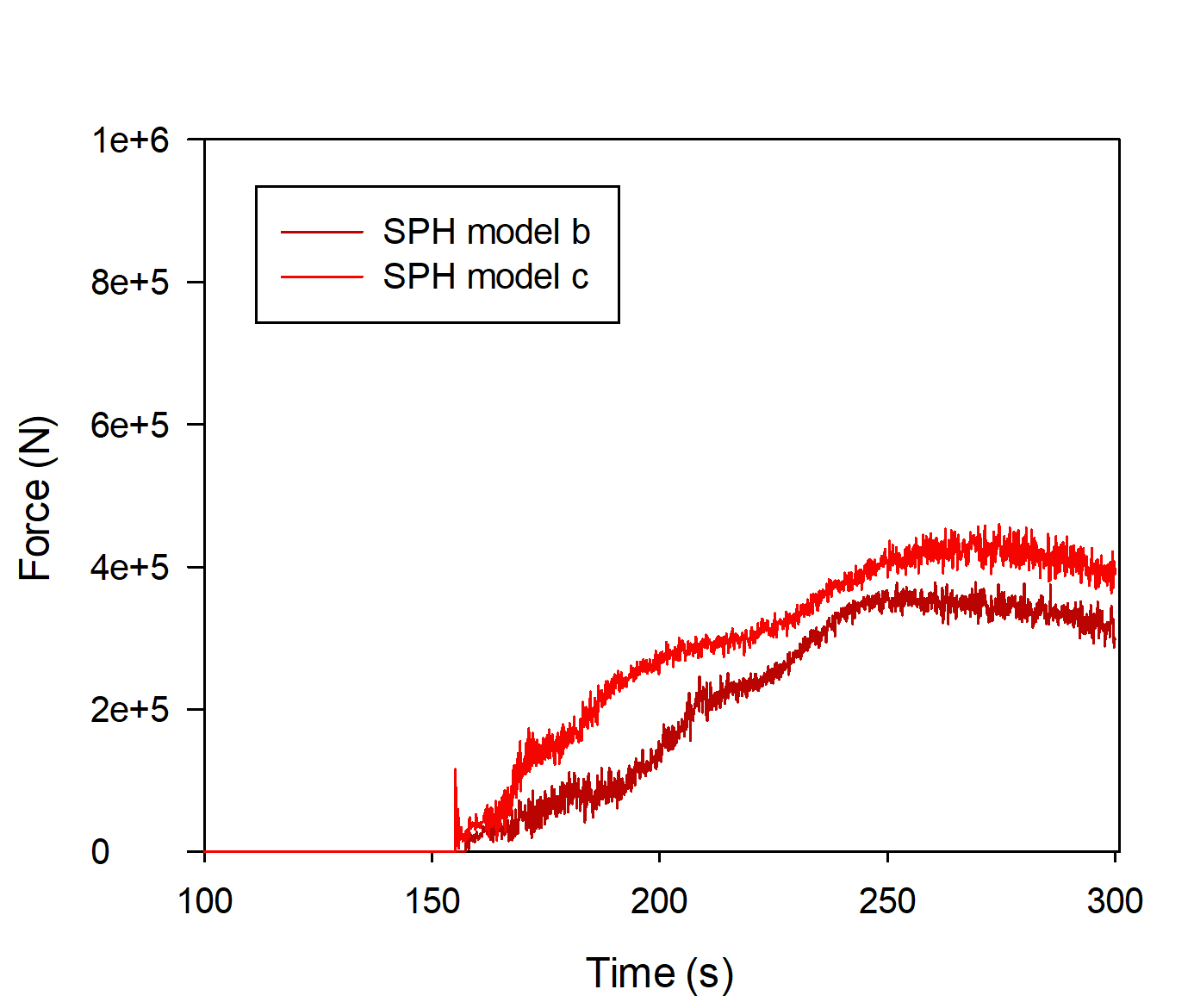}}\\
       	\subfloat[]{
       		\includegraphics[width=0.5\textwidth]{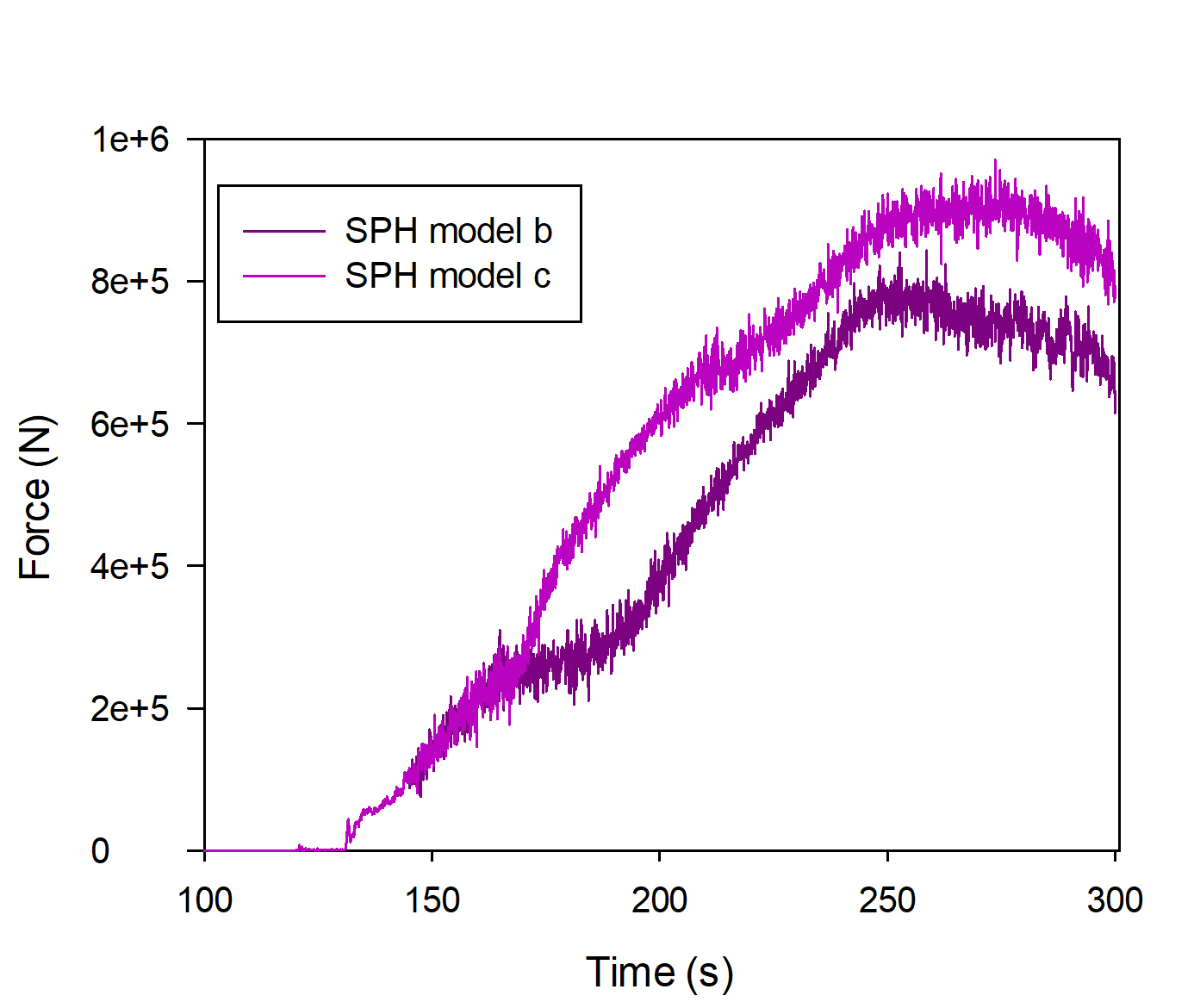}}
       	\subfloat[]{
       		\includegraphics[width=0.5\textwidth]{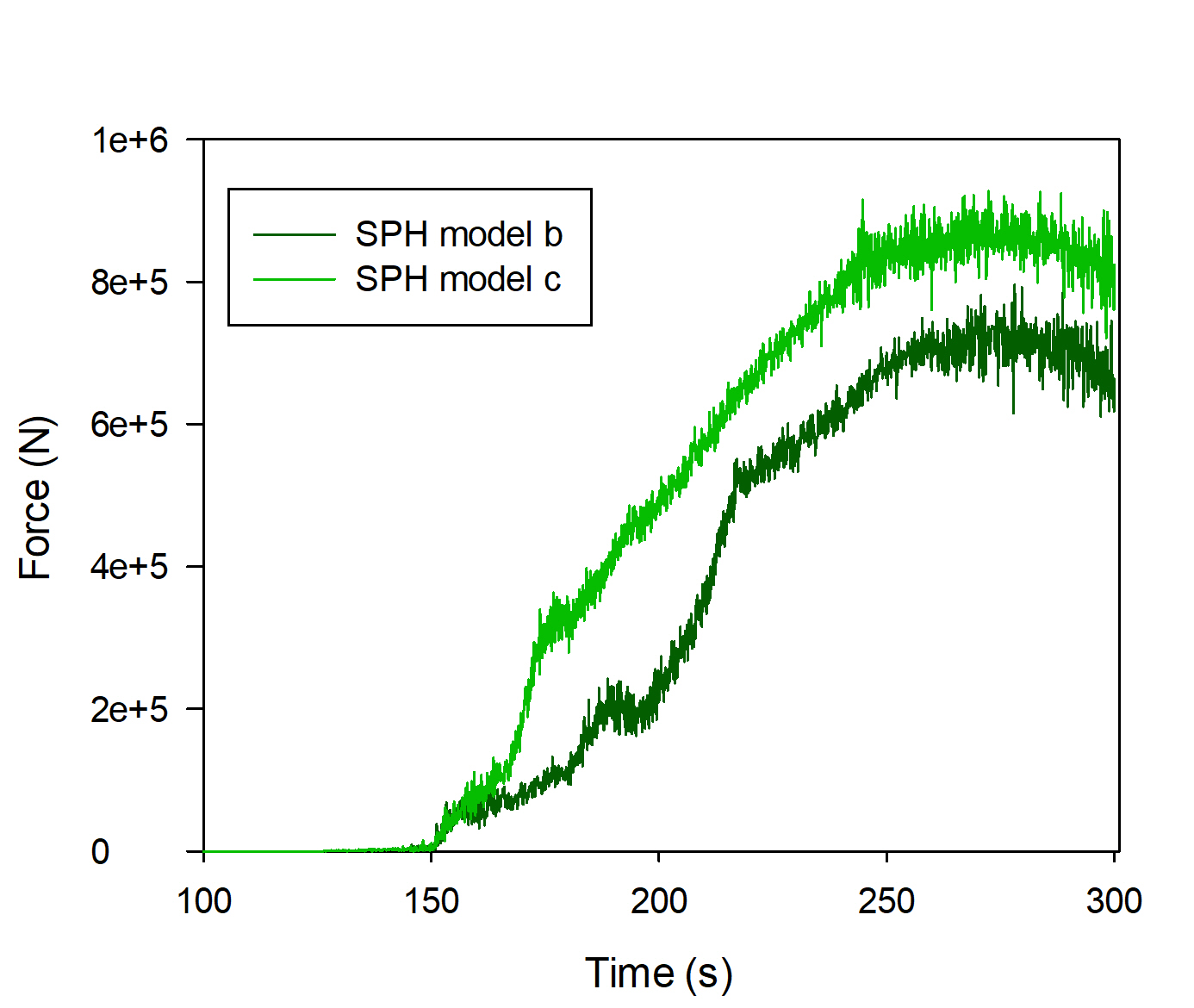}}
       	\caption{Variation of the hydrodynamic forces on building A; (a) \textit{Front Force}, (b) \textit{Back Force}, (c) \textit{Left Force} and (d) \textit{Right Force}.}
       	\label{fig:ferrea}
       \end{figure*}

       Figure \ref{fig:ferrea} shows the comparison of the forces on each wall: front (Figure \ref{fig:ferrea}.a), back (Figure \ref{fig:ferrea}.b), left (Figure \ref{fig:ferrea}.c) and right (Figure \ref{fig:ferrea}.d). Forces calculated in the SPH model c are higher in all the walls of building A.
       This indicates that a better resolution of the model topography produces an increase in the value of forces due to tsunamis that act on the building.
       
       \section{Conclusions}
       
       For this work we report the calculation of the hydrodynamic force generated by the tsunami on a building using a SPH real-scale simulation and a comparison with the classical approach provided by the Chilean standard NCh3363. It was observed that the temporal evolution of the hydrodynamic force from the SPH simulation is in good agreement with the classical approach. However, the maximum force computed by the SPH model is 30\% greater than the one computed by the Chilean standard, which may be due to the fact that the latter is based on hydraulic variables that do not consider the presence of the building, which implies an underestimation of the real impact force. It was also observed that the classical approach gave a greater impact force than the maximum hydrodynamic force estimated by the SPH model, obtaining a greater safety factor, and probably an overestimation of forces. 
       
       The comparison between simulations with different topography resolutions demonstrates that if the railway is not well represented, forces acting on the building could be underestimated. This highlights the importance of the use of high resolution topographic details in the SPH simulations. 
       
       This work has two important implications for the case study. The first is that the empirical estimation could overestimate the impact force, this shows that only through models that include the shock of the flow on the structure can the impact force be recreated. 
       The second is that the implementation of topographic models of higher resolution could imply unfavorable conditions in estimating forces, as is the case in this work. 
       
       Even though human safety is the main aspect to consider greater safety factors for structural design of buildings, economics requires efficient designs that are not structurally overestimated. This highlights the importance of a better estimate of forces of different nature on buildings. Finally, this work opens the way in real scale SPH tsunami simulations, considering real topographic conditions and databases of tides and buoys.
       \\\\
       \textbf{Acknowledgements:} This work was financed by the \textit{Ibero-American Programme for the Development of Science and Technology} (CYTED) under the Project 516RT0512 and by CINVESTAV-ABACUS (CONACyT grant EDOMEX-2011-C01-165873). The numerical calculations in this paper made use of the ABACUS-I supercomputing of the Centro de Matematica Aplicada y Computo de Alto Rendimiento, CINVESTAV-ABACUS. We thank the EPHYSLAB group of the University of Vigo (Spain), creators of the DualSPHysics code. The Authors also thank the CONICYT (Chile) for the grant FONDAP 15110017.


\begin{thebibliography}{99}
 	        
       	\bibitem{ara16} Ar\'anguiz R, Gonz\'alez G, Gonz\'alez J  et al (2016) The 16 September 2015 Chile Tsunami from the Post-Tsunami Survey and Numérical Modeling Perspectives. Pure and Applied Geophysics 173(2): 333-348.  
\smallskip
       	\bibitem{ara17} Ar\'anguiz R, Urra L, Okuwaki R, Yagi Y (2017) Development and application of tsunami fragility curve of the 2015 tsunami in Coquimbo, Chile. Nat. Hazards Earth Syst. Sci. https://doi.org/10.5194/nhess-2017-364 
\smallskip       	
       	\bibitem{asce} ASCE (2016) 2016, Minimum Design Loads for Buildings and Other Structures, ASCE/SEI Standard 7-05. American Society of Civil Engineers, Reston, Virginia. 
\smallskip        
       \bibitem{Carvajal16}  Carvajal, M., M. Cisternas, A. Gubler, P. A. Catalán, P. Winckler, and R. L. Wesson (2016), Reexamination of the magnitudes for the 1906 and 1922 Chilean earthquakes using Japanese tsunami amplitudes: Implications for source depth constraints, J. Geophys. Res. Solid Earth, 122, doi:10.1002/2016JB013269 
\smallskip      	
       	\bibitem{cre15} Crespo AJ, Dom\'inguez JM, Rogers BD, G\'omez-Gesteira M, Longshaw S, Canelas R et al (2015) DualSPHysics: Open-Source parallel CFD solver based on Smoothed Particle Hydrodiynamics (SPH). Computer Physics Communications 187: 204-216.
\smallskip       	
       	\bibitem{cre07} Crespo AJ, G\'omez-Gesteira M,  Dalrymple RA (2007) 3D SPH simulation of large waves mitigation with a dike. Journal of Hydraulic Research 45(5): 631-642.
\smallskip       	
       	\bibitem{deleffe} De Leffe M, Le Touz\'e D,  Alessandrini B (2010) SPH modeling of shallow-water coastal flows. Journal of Hydraulic Research 48(S1): 118-125.
\smallskip       	
       	\bibitem{mon77} Gingold RA, Monaghan JJ (1977) Smoothed particle hydrodynamics: theory and application to non- spherical stars. Monthly notices of the royal astronomical society 181(3): 375-389.
\smallskip
       	\bibitem{goncao2} Gonz\'alez-Cao J, Garc\'ia-Feal O, Dom\'inguez JM, Crespo AJ, G\'omez-Gesteira M (2018) Analysis of the hydrological safety of dams combining two numerical tools: Iber and DualSPHysics. Journal of Hydrodynamics 30(1): 87-94.		
\smallskip
       	\bibitem{FEMA} Heintz JA, Mahoney M (2008) Guidelines for Design of Structures for Vertical Evacuation from Tsunamis. FEMA P646/June 2008.
\smallskip       	
\bibitem{monag92}
Monaghan JJ (1992) Smoothed particle hydrodynamics. Annual review of astronomy and astrophysics, 30(1), 543-574.
\smallskip  
       	\bibitem{onemi15} ONEMI (2015) Oficina Nacional de Emergencia del Ministerio del Interior y Seguridad P\'ublica. Monitoreo por sismo de mayor intensidad. Publishing ONEMI Web. http://www.onemi.cl/alerta/se-declara- alerta-roja-por-sismo-de-mayor-intensidad-y-alarma-de-tsunami/ 
\smallskip       	
       	\bibitem{shoa0} SHOA (2000) Servicio Hidrogr\'afico y Oceanogr\'afico de la Armada de Chile. C\'omo sobrevivir a un maremoto. Valpara\'iso, Chile. Primera edici\'on, 15 p.
\smallskip       	
\bibitem{shoa2015} SHOA (2015) Carta de Inundaci\'on por tsunami, La Serena-Coquimbo. Servicio Hidrogr\'afico y Oceanogr\'afico de la Armada de Chile. available at     http://www.shoa.cl/servicios/citsu/pdf/CITSU\_ Coquimbo\_La\_Serena\_2da\_Ed.\_2015.pdf
\smallskip        
       	\bibitem{Soloviev84} Soloviev SL, Go CN (1984) Catalogue of tsunamis on the eastern shore of the Pacific Ocean. Institute of Ocean Sciences, Dept. of Fisheries and Oceans.
\smallskip      
       	\bibitem{StGermain13} St-Germain P, Nistor I, Townsend R \& Shibayama T (2013) Catalogue of tsunamis on the eastern shore of the Pacific Ocean. Institute of Ocean Sciences, Dept. of Fisheries and Oceans.
\smallskip     
       	\bibitem{wei} Wei Z, Dalrymple RA (2016)
       	Numerical study on mitigating tsunami force on bridges by an SPH model. Journal of Ocean Engineering and Marine Energy 2(3): 365-380.      
\smallskip       	
       	\bibitem{wei2} Wei Z, Dalrymple RA, H\'erault A, Bilotta G, Rustico E, Yeh H (2015) SPH modeling of dynamic impact of tsunami bore on bridge piers. Coastal Engineering 104: 26-42.   
\smallskip            
\bibitem{yam09}        
Yamazaki Y, Kowalik Z \& Cheung KF (2009) Depth‐integrated, non‐hydrostatic model for wave breaking and run‐up. International journal for numerical methods in fluids, 61(5), 473-497.
\smallskip        
       	\bibitem{yami11} Yamazaki Y, Cheung KF, Kowalik Z (2011) Depth-integrated, non-hydrostatic model with grid nesting for tsunami generation, propagation, and run-up. International Journal for Numerical Methods in Fluids 67(12): 2081-2107.
       	      	
       \end{thebibliography}
    \end{document}